\documentclass[aps,prd,twocolumn,superscriptaddress]{revtex4-1}

\usepackage{graphicx}
\usepackage{dcolumn}
\usepackage{bm}
\usepackage[caption=false]{subfig}

\begin{document}


\title{Supernova Relic Neutrino Search at Super-Kamiokande}



\affiliation{Kamioka Observatory, Institute for Cosmic Ray Research, University of Tokyo, Kamioka, Gifu, 506-1205, Japan}
\affiliation{Research Center for Cosmic Neutrinos, Institute for Cosmic Ray Research, University of Tokyo, Kashiwa, Chiba 277-8582, Japan}
\affiliation{Institute for the Physics and Mathematics of the Universe, Todai Institutes for Advanced Study, University of Tokyo, Kashiwa, Chiba 277-8582, Japan}
\affiliation{Department of Theoretical Physics, University Autonoma Madrid, 28049 Madrid, Spain}
\affiliation{Department of Physics, Boston University, Boston, Massachusetts 02215, USA}
\affiliation{Department of Physics and Astronomy, University of California, Irvine, Irvine, California 92697-4575, USA}
\affiliation{Department of Physics, California State University, Dominguez Hills, Carson, California 90747, USA}
\affiliation{Department of Physics, Chonnam National University, Kwangju 500-757, Korea}
\affiliation{Department of Physics, Duke University, Durham North Carolina 27708, USA}
\affiliation{Junior College, Fukuoka Institute of Technology, Fukuoka, Fukuoka 811-0214, Japan}
\affiliation{Department of Physics, Gifu University, Gifu, Gifu 501-1193, Japan}
\affiliation{Department of Physics and Astronomy, University of Hawaii, Honolulu, Hawaii 96822, USA}
\affiliation{High Energy Accelerator Research Organization (KEK), Tsukuba, Ibaraki 305-0801, Japan}
\affiliation{Department of Physics, Kobe University, Kobe, Hyogo 657-8501, Japan}
\affiliation{Department of Physics, Kyoto University, Kyoto, Kyoto 606-8502, Japan}
\affiliation{Department of Physics, Miyagi University of Education, Sendai, Miyagi 980-0845, Japan}
\affiliation{Solar Terrestrial Environment Laboratory, Nagoya University, Nagoya, Aichi 464-8602, Japan}
\affiliation{Department of Physics and Astronomy, State University of New York, Stony Brook, New York 11794-3800, USA}
\affiliation{Department of Physics, Okayama University, Okayama, Okayama 700-8530, Japan}
\affiliation{Department of Physics, Osaka University, Toyonaka, Osaka 560-0043, Japan}
\affiliation{Department of Physics, Seoul National University, Seoul 151-742, Korea}
\affiliation{Department of Informatics in Social Welfare, Shizuoka University of Welfare, Yaizu, Shizuoka, 425-8611, Japan}
\affiliation{Department of Physics, Sungkyunkwan University, Suwon 440-746, Korea}
\affiliation{Department of Physics, Tokai University, Hiratsuka, Kanagawa 259-1292, Japan}
\affiliation{The University of Tokyo, Bunkyo, Tokyo 113-0033, Japan}
\affiliation{Department of Engineering Physics, Tsinghua University, Beijing, 100084, China}
\affiliation{Institute of Experimental Physics, Warsaw University, 00-681 Warsaw, Poland}
\affiliation{Department of Physics, University of Washington, Seattle, Washington 98195-1560, USA}

\author{K. Bays}

\affiliation{Department of Physics and Astronomy, University of California, Irvine, Irvine, California 92697-4575, USA}

\author{T. Iida}
\author{K. Abe}

\affiliation{Kamioka Observatory, Institute for Cosmic Ray Research, University of Tokyo, Kamioka, Gifu, 506-1205, Japan}

\author{Y. Hayato}

\affiliation{Kamioka Observatory, Institute for Cosmic Ray Research, University of Tokyo, Kamioka, Gifu, 506-1205, Japan}
\affiliation{Institute for the Physics and Mathematics of the Universe, Todai Institutes for Advanced Study, University of Tokyo, Kashiwa, Chiba 277-8582, Japan}

\author{K. Iyogi}
\author{J. Kameda}
\author{Y. Koshio}
\author{L. Marti}
\author{M. Miura}

\affiliation{Kamioka Observatory, Institute for Cosmic Ray Research, University of Tokyo, Kamioka, Gifu, 506-1205, Japan}

\author{S. Moriyama}
\author{M. Nakahata}

\affiliation{Kamioka Observatory, Institute for Cosmic Ray Research, University of Tokyo, Kamioka, Gifu, 506-1205, Japan}
\affiliation{Institute for the Physics and Mathematics of the Universe, Todai Institutes for Advanced Study, University of Tokyo, Kashiwa, Chiba 277-8582, Japan}

% ...

\collaboration{The Super-Kamiokande Collaboration}
\noaffiliation

\date{\today}

\begin{abstract}
A new Super-Kamiokande (SK) search for Supernova Relic Neutrinos (SRNs) was conducted using 2853 live days of data.  Sensitivity is now greatly improved compared to the 2003 SK result, which placed a flux limit near many theoretical predictions.  This more detailed analysis includes a variety of improvements such as increased efficiency, a lower energy threshold, and an expanded data set.  New combined upper limits on SRN flux are between 2.8 and 3.0 $\bar{\nu}_e$ cm$^{-2}$ s$^{-1}$ $>$ 16 MeV total positron energy (17.3 MeV $\mbox{E}_\nu$).  
\end{abstract}

\pacs{}

\maketitle

\section{Introduction and Motivation}
\subsection{The SRN signal}

	With supernovae each releasing on the order of $10^{46}$ J of energy, 99$\%$ as neutrinos, the neutrino flux from each supernova event is enormous.  In a galaxy such as the Milky Way, it is estimated a supernova will occur around 2 or 3 times a century \cite{snrate}.  Although we would not see a significant neutrino burst from a supernova farther away than our galaxy and its satellites, all of the core collapse supernovae that have exploded throughout the history of the universe have released neutrinos, which, in the absence of unexpected physics, should still be in existence.  These neutrinos are herein called the supernova relic neutrino (SRN) signal (or simply relics, though care should be taken, since this terminology is also sometimes used for big bang relic neutrinos), also sometimes referred to as the Diffuse Supernova Neutrino Background (DSNB).

\begin{figure}[!t]
\includegraphics[width=3in]{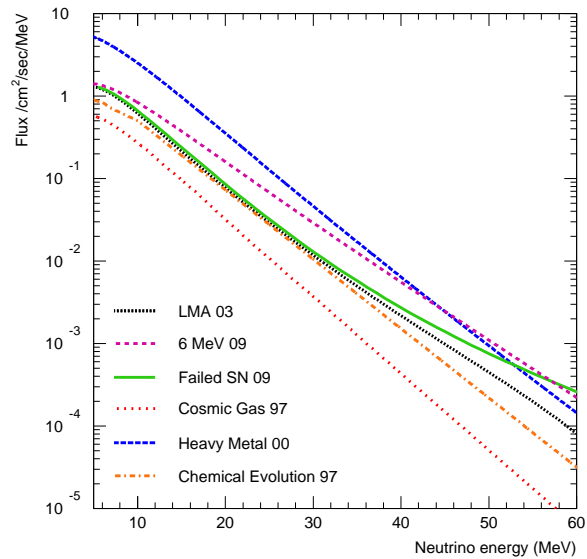}
\caption{Examples of theoretical SRN spectra.  Flux is $\bar{\nu}_e$ only.  For heavy metal abundance model, average metal yield of one solar mass is assumed.}
\label{srnspec}
\end{figure}

	Many astrophysicists have considered the SRN signal, constructing models that predict both the flux and the spectrum.  The first models were crafted even before SN1987A \cite{bisno,krauss,woosley}, then after SN1987A with increasing interest and sophistication, with models such as: Totani \textit{et al.}'s constant SN rate model \cite{totani1996}; Malaney's cosmic gas infall model \cite{malaney}; Woosley and Hartmann's chemical evolution model \cite{hartmann}; Kaplinghat \textit{et al.}'s heavy metal abundance model \cite{kap2000}; Ando \textit{et al.}'s LMA model \cite{ando2003} \footnote{The flux of the LMA model is increased by a factor of 2.56 from the paper, a revision introduced at NNN05}; Lunardini's failed supernova model \cite{lunardini} \footnote{Assumed parameters are: Failed SN rate = 22$\%$, EoS = Lattimer-Swesty, and survival probability = 68$\%$}; and the variable neutrino temperature formulation of Horiuchi \textit{et al.} (6 MeV and sometimes 4 MeV cases considered) \cite{isdetect}.  Some examples of theoretical spectra are shown in Fig. \ref{srnspec}.  Although the normalizations vary, the general shape and slope of the predicted spectra are relatively similar.  

	The SRN signal has never been seen.  A paper was published in 2003 detailing the first search for the SRN events at SK \cite{malekpaper}.  The basic method was to eliminate as many backgrounds as possible, then attempt to model the spectrum of the remaining backgrounds.  A $\chi^{2}$ fit was performed on the energy spectrum of the final data sample, with two background components and one signal component.  From this fit, a final model independent flux limit of 1.2 $\bar{\nu}_e$ cm$^{-2}$ s$^{-1}$ for $\bar{\nu}_e$ energy $>$ 19.3 MeV was extracted, which was $\sim$100 times more stringent than the previous world's best limit \cite{srn88}.  This study is an update to that 2003 result, with more livetime, lower energy threshold, and significantly increased sensitivity.  Other experiments have also produced SRN limits, such as SNO \cite{snorelic} and KamLAND \cite{kamdsnb}.  Recent models predict a SRN flux that is on the cusp of discovery \cite{strigkaprelic,isdetect}, motivating further efforts.  

\subsection{The SK detector}

	The Super-Kamiokande detector is a 50 kton water Cherenkov detector located in the Kamioka mine in Japan.  The inner detector (32.5 ktons, 22.5 ktons fiducial) has an inner surface lined with $\sim$11,100 ($\sim$5,200 for SK-II) 50 cm Hamamatsu PMTs.  More information on the hardware and experimental setup of the detector, including calibration details, is provided by the following references: \cite{sknim,linacpaper,n16paper}.
	
	The SK detector began data taking in April 1996, and was shut down for maintainance in July 2001.  This period is referred to as SK-I (1497 days livetime).  While refilling after the maintainance, on November 12, 2001, a major accident \cite{accident} destroyed 60$\%$ of the PMT tubes.  The surviving tubes were redistributed to cover the entire detector while replacement tubes were manufactured, and data taking was resumed in December 2002, running until October 2005 with reduced cathode coverage.  This is the SK-II period (794 days).  After the replacement tubes were installed, and the detector was back up to its nominal 40$\%$ coverage, data taking again continued from July 2006 until August 2008, marking the SK-III period (562 days).  Finally, the detector electronics were upgraded \cite{newdaq,newdaq2}, and data taking since September 2008 is referred to as SK-IV.  This study considers SK-I, SK-II, and SK-III data.

	The SK detector has many different triggers.  When enough PMTs observe light within 200 ns (approximately the longest time it can take light to traverse the inner detector) a detector trigger occurs.  Inner detector (ID) electronics triggers include the High Energy (HE) trigger (about 33 PMTs), Low Energy (LE) trigger (about 29 PMTs, 100$\%$ efficient at 6.5 MeV), and Super Low Energy (SLE) triggers (variable threshold between 17 and 24 PMTs).  SLE events have the lowest threshold (which has changed over time), and occur orders of magnitude more often than LE triggered events ($\sim$ 1 KHz as opposed to $\sim$ 10 Hz), requiring dedicated hardware for filtering and reducing the SLE data flow to manageable levels.  The SLE trigger enables SK to see the lowest energy events.  The outer detector (OD) also triggers if 19 or more OD PMTs fire within 200 ns.  

	SK primarily sees SRN events via inverse beta decay ($\bar{\nu}_e + p \rightarrow n + e^+$).  The next most visible mode for SRN interactions is about two orders of magnitude less frequent.  The inverse beta decay cross section used is described in references \cite{strumia,beacomvogel}.

\section{Data reduction}
	
\subsection{Cuts}

\setcounter{secnumdepth}{1} 

	Each SK PMT fires with a dark rate of $\sim$3-5 kHz; cosmic ray muons penetrate the detector at $\sim$2 Hz, and $\sim$25 atmospheric and solar neutrino events are identified every day.  Many of these act as backgrounds to the SRN signal.  With an expected rate of only a couple of SRN events a year, careful background reduction is necessary.  Most backgrounds are removable, but a few remain after all cuts and must be understood.  The events that can be removed, and the cuts used to remove them, are described in this section.	
	
	The software tools used for reconstructing event vertex, direction, and energy information are the same as used in the long-standing SK solar neutrino analysis (\cite{sk1solar,sk2solar}).

\subsection{Noise reduction}

	Many noise events can be separated from the main sample easily.  These preliminary cuts are the same as those used for the long-studied SK solar analysis \cite{sk1solar}, and are collectively called `noise reduction'. 
	
	Calibration events are labeled as such and are immediately removed.  Events which trigger the OD are also removed, as these events have a charged particle entering the detector from outside, while relic events should be fully contained.  Electronics noise events are also cut using timing and charge considerations, with a signal efficiency of almost 100$\%$.

	The total charge deposited in the detector by the event is a quick criterion useful for separating out muons.  Muons are in general much brighter than inverse beta decay positrons, and deposit much more charge in the detector.  Any event with more than 800 photo electrons is removed (400 p.e. for SK-II), quickly eliminating most of the muons.  

\subsection{Fiducial volume cut}

	The SK ID is filled with 32.5 ktons of pure water.  Radioactive events occur in the detector from sources such as the surrounding rock wall and radioactive isotopes in the PMT glass itself.  Most of these events are eliminated by definition of a fiducial volume (FV).  In the relic analysis, we have implemented the standard SK FV cut of removing the volume within 2 meters of the walls, leaving a FV mass of 22.5 ktons.	 
	
\subsection{Spallation cut}

	The analysis lower energy threshold is determined by spallation, which is the process of a nucleus emitting nucleons when struck by a highly energetic particle.  Cosmic ray muons interacting with an oxygen nucleus cause radioactive spallation which mimics a real SRN signal.  At lower energies, the unstable nuclei tend to have longer lifetimes, making them harder to tag through muon correlation.  Eventually tagging becomes too inefficient for the remaining data to be useful, thus determining the lower energy threshold of the analysis.  Spallation constitutes the bulk of the SRN background after the noise reduction cut. Elimination of spallation events is complex and quite costly in terms of signal efficiency, making it perhaps the most important removable background.

	To remove spallation, we correlate relic candidates to preceding muons.  If the muon is responsible for the event, then the event should be located near the muon track, both in location and time.  The main spallation cut utilizes a likelihood method:
	
	First, the muon tracks are reconstructed and categorized as: (i) single muons that travel through the detector ($\sim83\%$ of muon events); (ii) multiple muons bundles ($\sim8\%$); (iii) muons that enter from outside but stop within the ID (stopping muons, $\sim5\%$); and (iv) short track length muons, usually just clipping the corner of the detector ($\sim4\%$).  Each of these muon types was found to produce spallation differently, and are thus treated independently.
 
	A dE/dx profile of each muon track is constructed by projecting back the amount of light seen by each PMT and determining where along the muon track it originated according to the measured arrival time of the light at the PMT.  This profile of charge emitted along the muon track tends to peak for spallation producing muons approximately at the position along the track the spallation occurred.  Although the exact mechanism responsible for this peak is not known, the correlation is unmistakable in our data. 
 	
\begin{figure}[!t]
\centering{
\includegraphics[width=2in]{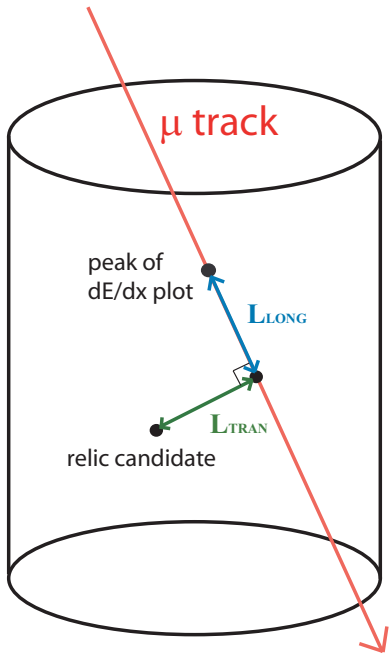}}
\caption{Schematic explanation of spallation distance variables.}
\label{schem}
\end{figure}

 		 Four spallation tagging variables are considered.  The first spallation variable, $\Delta$t, is simply the amount of time by which the muon preceded the relic candidate.  The second variable, $L_{\mbox{\scriptsize{TRANS}}}$, is the transverse, or perpendicular, distance from the muon track to the relic candidate (see Fig. \ref{schem}).  The center of the nine consecutive bins (4.5 m distance) with the largest combined charge in the dE/dx histogram corresponds to a position along the muon track where the spallation is expected to have occurred.  The distance along the muon track from where the spallation is expected to occur to where the relic candidate exists is the third likelihood variable, longitudinal distance $L_{\mbox{\scriptsize{LONG}}}$.  Lastly, the value of the combined charge in the largest nine consecutive bins of the dE/dx histogram is our fourth likelihood variable, $Q_{\mbox{\scriptsize{PEAK}}}$.  The more charge in the peak, the more likely the muon is to be making spallation.
 	
\begin{figure}[!t]
\hspace{0.8mm}
\includegraphics[width=3.1in]{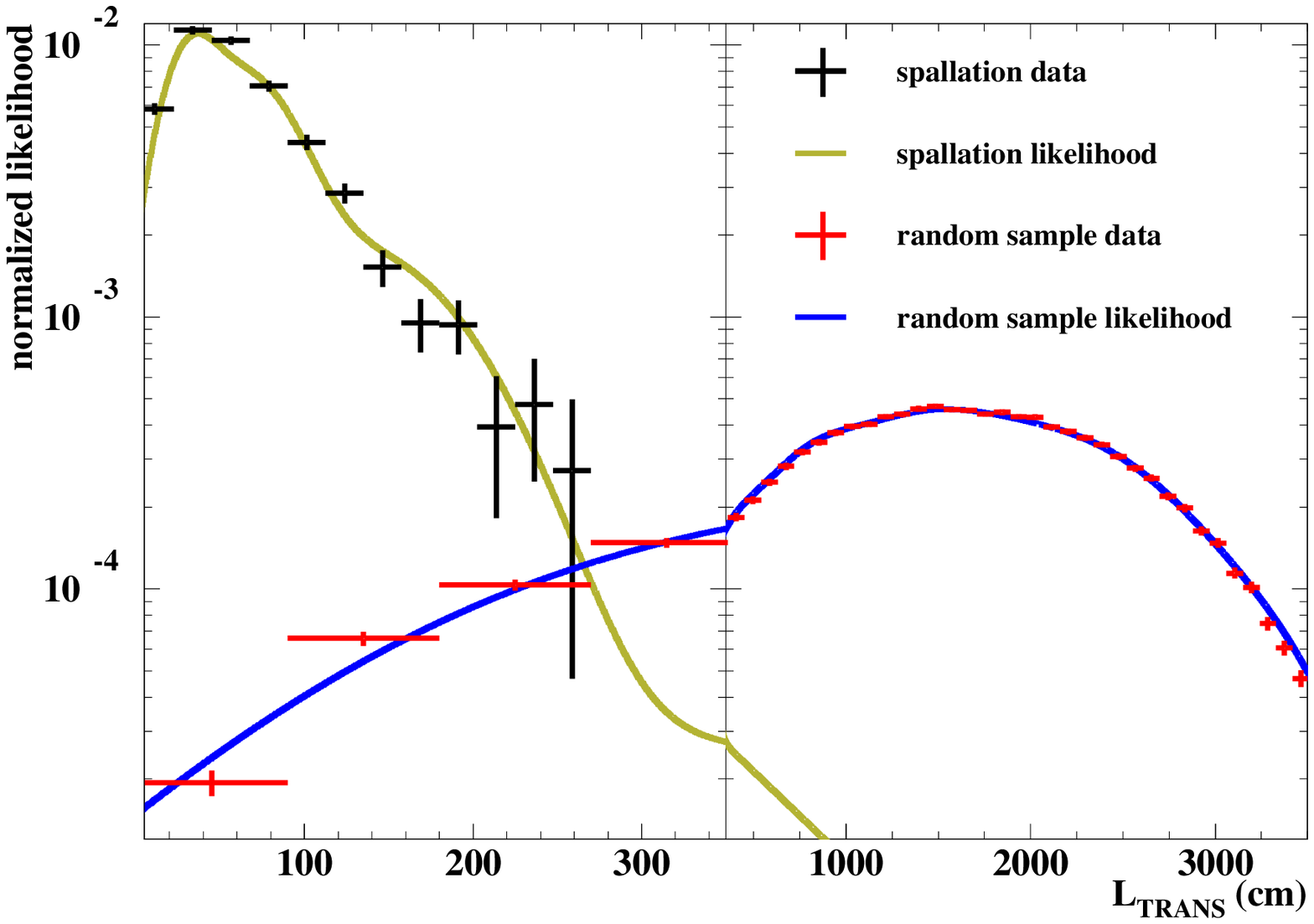}
\includegraphics[width=3.2in]{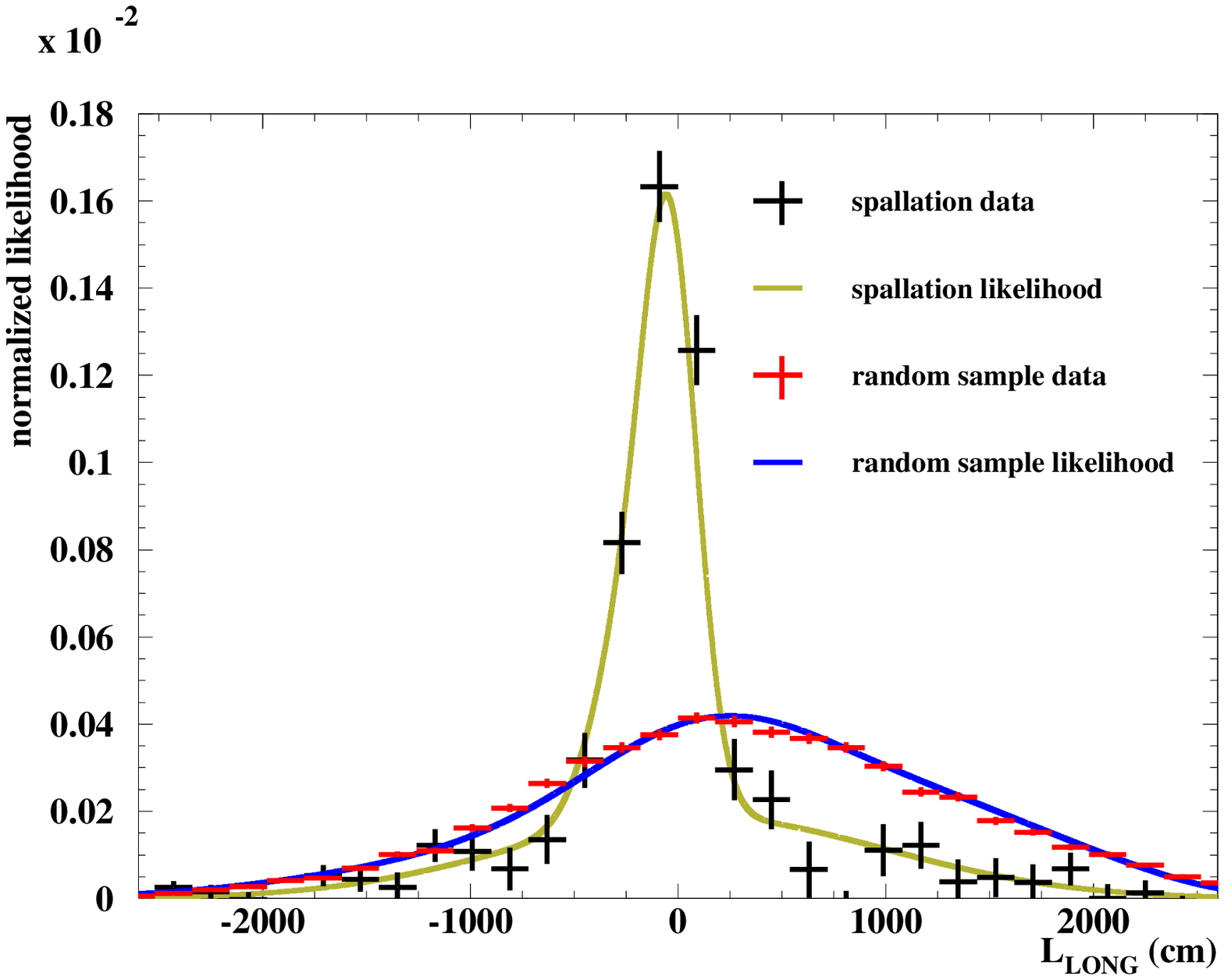}
\caption{SK-I/III data with likelihood functions overlaid for single through-going muons.  Top shows transverse distance, bottom shows longitudinal distance.}
\label{spalllt}
\end{figure}

	 Probability density functions (PDFs) are formed for each of the four spallation variables.  For each muon categorization, the relic candidates were correlated to muons preceding the candidate in time (the `data sample'), and correlations were examined.  A random sample was formed by taking the same correlations to muons immediately following the relic candidate in time.  The random sample histograms were subtracted from the data histograms (yielding a `spallation sample'), for each muon categorization.  These profiles were parameterized, resulting in functions representing spallation (from the spallation sample) and accidental correlation background (from the random sample).  These parameterizations, once normalized, are the PDFs, which are multiplied together to give the likelihood (for example, see Fig. \ref{spalllt}).

\begin{figure}[!t]
\centering
\includegraphics[height=2.4in]{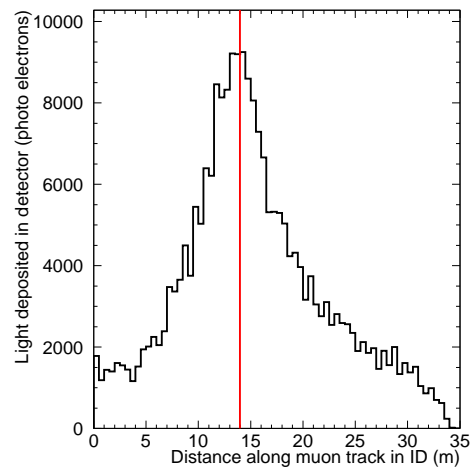}
\caption{Example of a dE/dx plot.  The red line indicates where along the muon track the candidate was reconstructed.  This example has particularly good correlation.}
\label{dedx}
\end{figure}	  
	 
		For each muon categorization, a cut on the likelihood was instituted.  Cut values were tuned until no statistically significant difference existed in the distributions of the data remaining after the cut compared to the random sample for the $\Delta$t and $L_{\mbox{\scriptsize{TRANS}}}$ variables.  The spallation contamination remaining after the spallation cut is difficult to estimate due to the large statistical uncertainties, but no evidence for remaining background was found.  For the SK-I/III combined sample, we could see any excess with a resolution of about 4 events; and for SK-II, with a resolution of about 2 events.
		
\begin{figure}[!t]
\centering
\includegraphics[width=3in]{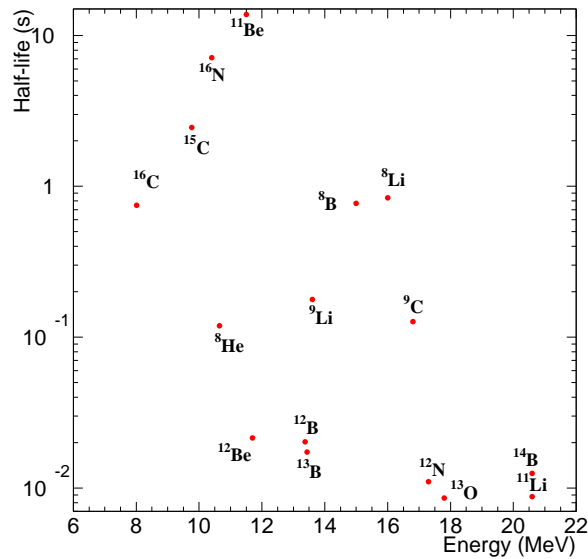}
\caption{Half-lives and event end-point energies of spallation products expected to occur in pure water.  Lower energy events trend to longer decay times.}
\label{scatterspall}
\end{figure}

		SK-I and SK-III use the same spallation cut.  Due to the differences in cathode coverage, SK-II required separate likelihoods and tuning.  The lower energy threshold of the cut is the energy below which keeping the sample free of any definite spallation contamination becomes too inefficient to be viable.  Fig. \ref{scatterspall} shows the half-life and maximum energy of spallation products in SK; as spallation products decrease in energy, they tend to have longer half-lives, and are more difficult to tag.  Taking into account energy resolution, events at 14 MeV such as $^{9}\mbox{Li}$ can reconstruct up to 16 MeV.  The final energy threshold of the analysis, for which no evidence of spallation remains in the final sample, is 16 MeV positron energy (or 17.3 MeV $\bar{\nu}_e$ energy) for SK-I/III, and 17.5 MeV for SK-II.  

	The cut need not be applied at all energies.  The cut's upper energy bound is determined by the highest energy spallation products, $^{11}\mbox{Li}$ and $^{14}\mbox{B}$, each with an end point energy of 20.6 MeV and a half-life of around 0.01 seconds.  The SK-I/III spallation cut is applied up to 24 MeV positron energy, and the SK-II cut up to 26 MeV, which is about 1.5 $\sigma$ of energy resolution away from 20.6 MeV. 

	The cut is split into two energy regions: a more stringent selection for the lower energies, which has many more spallation events and is most critical for the analysis; and a looser selection at the higher energies, which has fewer spallation events, many of which are $^{11}\mbox{Li}$ and are easily removed considering its short half-life.  Thus, for SK-I/III, the cut from 16-18 MeV is different than from 18-24 MeV, and in SK-II, the cut from 17.5-20 MeV is different than from 20-26 MeV.

	For all muons, a timing cut rejecting events with $\delta$t $<$ 4 seconds (8 s) was implemented on the most energetic of muons, those over 400,000 (800,000) p.e., as these events are rare and almost always create spallation.  
	
	For multiple muon type events, the secondary track is usually well fit, so the dE/dx profile of secondary tracks was calculated, and a separate secondary track likelihood tuned.  Additional tracks do not have trustworthy dE/dx profiles, and a likelihood for these cases was constructed based only on $\Delta$t, $L_{\mbox{\scriptsize{TRANS}}}$, and the total amount of light deposited by the muon in the detector.  
	
		Finally, for single through-going muons, $\sim1.5\%$ of the time the muon fitter was unable to correctly fit the track.  A dedicated alternate fitter was developed for these misfit events, which correctly fit approximately 75$\%$ of the misfit single through-going muons.  For the remaining 25$\%$ of misfits, a 2 second detector veto was implemented to prevent spallation contamination.  
	
\begin{figure}[!t]
\centering
\includegraphics[width=2.5in]{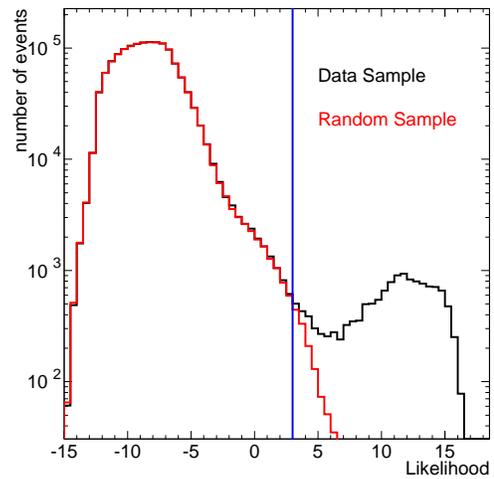}
\caption{Comparison of likelihood distributions of data sample and random sample for single through-going muons in SK-I/III.  The data sample contains spallation events, while the random sample does not.  The high likelihood values of the spallation excess show the effectiveness of the cut.  The SK-I/III 16-18 MeV cut value of 3 is shown; everything to the right of the line is rejected.}
\label{slplot}
\end{figure}

	Fig. \ref{slplot} demonstrates the power of this method to tag spallation events.  Table \ref{spaineff} shows the cut signal efficiency.  In the published 2003 SK analysis, in order to be spallation free the spallation cut was 37$\%$ inefficient from 18 to 34 MeV.  Now, for the same SK-I data, the cut is only 9$\%$ inefficient from 18 to 24 MeV, and furthermore the 16-18 MeV region is now opened up for study with an inefficiency of 18$\%$, all spallation free.  
		
\begin{table}
\caption[Spallation Cut Signal Efficiencies]{Spallation Cut Signal Efficiencies}
\label{spaineff}
\begin{center}
\begin{tabular}{|c|c|c|}
\hline
\textbf{EFFICIENCY} & \textbf{SK-I/III} & \textbf{SK-II}\\
\hline
Low energy & 81.8 $\%$ (16-18 MeV) & 76.2 $\%$(17.5-20 MeV)\\
High energy & 90.8 $\%$ (18-24 MeV) & 88.2 $\%$(20-26 MeV)\\
\hline
\end{tabular}
\end{center}
\end{table}

\subsection{Solar angle cut} 

	Solar neutrinos elastically scatter off electrons, creating recoil electrons that look like SN relic positrons.  To discriminate such events, we use the angle between the reconstructed direction vector of the event and the direction vector from where the sun was in the sky at the time of the event ($\theta_{\mbox{\tiny{sun}}}$).
	
	Although the direction of the recoil electron closely follows that of the incoming solar neutrino (usually within $\sim 10$ degrees), multiple Coulomb scattering (MCS) of the recoil electron smears out the peak in the cos($\theta_{\mbox{\tiny{sun}}}$) distribution at cos($\theta_{\mbox{\tiny{sun}}}$) $= 1$.  This long tail in the distribution makes cutting solar neutrino events using cos($\theta_{\mbox{\tiny{sun}}}$) inefficient, especially since the end points for $^{8}\mbox{B}$ and hep solar neutrinos are right at the critical lower energy edge of the analysis, where inefficiency impacts the relic sensitivity the most.  In order to maximize sensitivity, we used an energy dependent cut coupled with an estimate of MCS.
	
\begin{figure}[!t]
\centering
\includegraphics[width=2.8in]{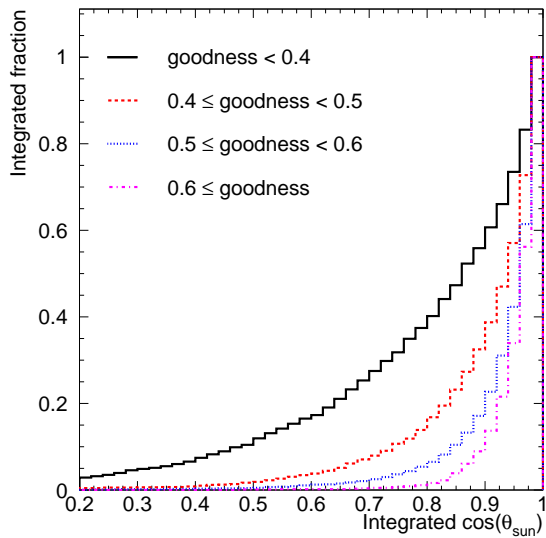}
\caption{SK-I/III solar MC integrated $\cos(\theta_{\mbox{\tiny{sun}}})$ distributions by multiple Coulomb scattering goodness bin.}
\label{papersolar}
\end{figure}

	MCS was estimated using a PMT hit by PMT hit Hough transform that reconstructs the direction and returns a goodness value corresponding to the amount of MCS.  The correlation of this MCS goodness to the shape of the cos($\theta_{\mbox{\tiny{sun}}}$) distribution can be seen in Fig. \ref{papersolar}.  The candidate sample is split into various MCS goodness bins, and the cut value for each MCS goodness bin is separately tuned, in one MeV energy bins, by use of the following significance function:
	
	\begin{equation}	
	\mbox{Significance} =  \epsilon / \sqrt{\kappa S + \epsilon \alpha}
	\end{equation}	
	
	where $\epsilon$ is the cut efficiency, $S$ is the number of solar neutrino events, $\kappa$ is the reduction effectiveness of the cut (such that $\kappa\times S$ is the number of solar neutrino events remaining after the cut is applied), and $\alpha$ represents the non-solar neutrino background.  
	
	The solar neutrino spectrum is modeled using Monte Carlo (MC).  The solar neutrino flux was normalized using real SK data below energy threshold (14-16 MeV), with all cuts applied (except the one on cos($\theta_{\mbox{\tiny{sun}}}$)).  To normalize, it was assumed that all events with cos($\theta_{\mbox{\tiny{sun}}}$) $<$ 0 were not solar neutrino events, and that all non-solar remaining events had no cos($\theta_{\mbox{\tiny{sun}}}$) dependence.  Thus, simply subtracting the number of events with cos($\theta_{\mbox{\tiny{sun}}}$) $<$ 0 from the number of events with cos($\theta_{\mbox{\tiny{sun}}}$) $>$ 0 allows us to estimate the number of solar neutrino events at a particular energy, and then extrapolate the number of solar neutrino events at all energies.  
	
	The dominant non-solar neutrino background was assumed to be decay electrons from atmospheric $\nu_{\mu}$ CC events, which is our largest background (see section III).  For the purpose of tuning the solar cut, the decay electron spectrum was modeled using a sample of data that was identified as decay electrons by timing and spatial correlation to a preceding muon that had no OD trigger (a fully contained muon from a muon neutrino interaction).  The normalization of the decay electron background was determined by fitting to the data.
	
	As with many of the cuts, SK-I and SK-III used one cut, and a separate version of the cut was tuned for SK-II.  The SK-II cut criteria is simply the SK-I/III cut with the energy bins shifted by $+6\%$ to reflect the poorer energy resolution.  The cut is MCS goodness bin dependent for E $<$ 19 MeV (SK-I/III, $<$ 20.14 MeV SK-II), while for 19-20 MeV (20.14-21.2 MeV SK-II) there is a simple, MCS goodness bin-independent cut that eliminates events with $\cos(\theta_{\mbox{\tiny{sun}}}) >$ 0.93.  More cut details can be found in Table \ref{solarsum1}.  Note that the cut values in the table mean that events with $\cos(\theta_{\mbox{\tiny{sun}}}) > $ cut value are rejected.  Also note that the inefficiencies are already weighted by the fraction of events in that MCS goodness bin, so that they can be simply added for the total.

	After the cut, no statistically significant evidence for remaining solar neutrino events exists, and the estimated number of remaining solar neutrino events in the combined SK-I/II/III final data sample is $<$ 2.

\begin{table}
\caption[SK-I/III solar cut inefficiency summary]{SK-I/III solar $\cos(\theta_{\mbox{\tiny{sun}}})$ cut value (inefficiency)}
\label{solarsum1}
\begin{center}
\begin{tabular}{|c|c|c|c|}
\hline
\textbf{MCS Goodness} & \textbf{16-17 MeV} & \textbf{17-18 MeV} & \textbf{18-19 MeV} \\ 
\hline
g $<$ 0.4          & 0.05 (7.2$\%$) & 0.35 (5.0$\%$) & 0.45 (4.3$\%$) \\
0.4 $\leq$ g $<$ 0.5 & 0.39 (9.7$\%$) & 0.61 (6.4$\%$) & 0.77 (3.8$\%$) \\
0.5 $\leq$ g $<$ 0.6 & 0.59 (6.7$\%$) & 0.73 (4.5$\%$) & 0.81 (3.2$\%$) \\
0.6 $\leq$ g       & 0.73 (2.4$\%$) & 0.79 (2.0$\%$) & 0.91 (1.0$\%$) \\
\hline
\textbf{Total} & 26.2$\%$ & 17.9$\%$ & 12.2$\%$ \\
\hline
\end{tabular}
\end{center}
\end{table}

\subsection{Incoming event cut}

	The FV cut eliminates most of the radioactivity events that enter from detector edges, but not all.  Instead of reducing the FV, it is more efficient to separate the remaining radioactive background by position and direction considerations.  
	
	We use an effective distance parameter, $d_{\mbox{\tiny{eff}}}$ which is defined identically as in the SK solar neutrino analysis (where this is referred to as the `Gamma Ray Cut' \cite{sk1solar}).  The event direction and position are reconstructed; then, starting at the event vertex, the distance projected backwards along the event direction is calculated until a wall is hit.  This distance is $d_{\mbox{\tiny{eff}}}$.  
	
	The radioactive backgrounds targeted by the incoming event cut are energy dependent, and much more prevalent at lower energies.  However, some background exists at all energies.  Thus, the cut has two pieces.  A 300 cm $d_{\mbox{\tiny{eff}}}$ cut is applied at all energies, for all three SK phases.  The lower energy part of the cut is tuned separately for each SK phase.
	
\begin{figure}[!t]
\centering
\includegraphics[width=2.8in]{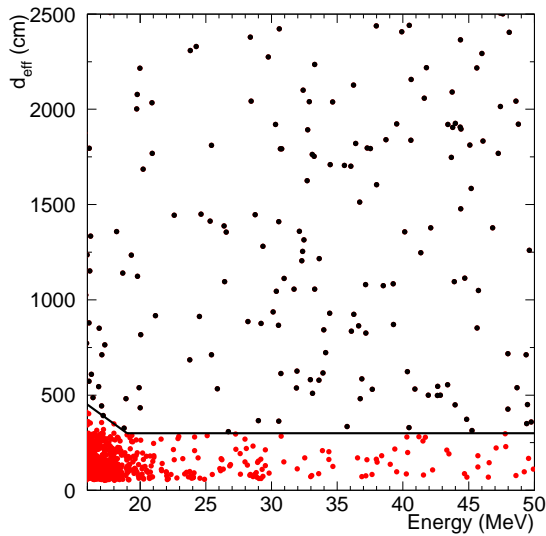}
\caption{SK-I data distribution showing energy dependent incoming event cut.  Events below the line are rejected.}
\label{effwall}
\end{figure}

	For SK-I, sufficient statistics existed to allow an energy dependent tuning of the cut, to maximize efficiency (see Fig. \ref{effwall}).  The shielding added to the PMTs at the start of SK-II to prevent another serious accident produce radioactive backgrounds, and thus the SK-I tuning could not be used for other phases.  Furthermore, the reduced cathode coverage of SK-II makes it different.  Each phase is tuned separately, but for SK-II and SK-III, the statistics were not sufficient for energy dependent tuning.  Instead, a cut was implemented below 22 MeV, at $d_{\mbox{\tiny{eff}}}$ $<$ 500 cm for SK-II, and $d_{\mbox{\tiny{eff}}}$ $<$ 450 cm for SK-III. 

\subsection{Decay electron cut}

	The SK electronics saves PMT signals in a 1.3 $\mu$s window.  Sometimes, a muon decays into an electron quickly enough that the light from both particles are captured in the same event.  These events are eliminated by searching for a multi-peak structure in the timing information of the event.
	
	More commonly, the muon and resultant decay electron are in separate events.  For each relic candidate, all SK data is searched up to 50 $\mu$s before the candidate (in case the candidate is a decay electron), and up to 50 $\mu$s after the event (in case the candidate is actually a low charge muon from an atmospheric neutrino interaction, or one of two decay electrons).  
	
	When searching for pre-activity, finding any LE or HE triggered event within the 50 $\mu$s causes the candidate to be rejected.  Since SLE events are much more frequent, a further vertex correlation requirement of 5 meters is required in order to improve efficiency.  When searching for post-activity, the 5 meter vertex correlation is required of all events, whether HE, LE, or SLE.
	
\subsection{Pion cut}

	Some of the higher energy events in our sample (usually $>$ 30 MeV, reconstructed assuming event is an electron) are charged pions from atmospheric neutrino interactions.  One way to distinguish between charged pions and electrons is by looking at the `fuzziness' of the resulting Cherenkov ring.  This is because pions are quickly captured by an oxygen nucleus, while electrons tend to travel farther and MCS more.  This MCS causes the electron direction to change as it travels, causing the Cherenkov ring to be more poorly defined.  Pions are also heavier, and deflect less when scattering.  
	
	Quantitatively, this difference can be used to construct a tool for discriminating pions and electrons.  When the Cherenkov angle for an event is calculated, a distribution of three-hit combinations (or triplets) is formed.  Each combination of three hit PMTs form a cone, and an opening angle.  The distribution of these opening angles has a peak that is narrower for pions than for electrons.  Using this logic, a pion ratio variable was constructed, where pion likelihood is defined as follows (using the triplet distribution):
	
	\begin{eqnarray}	
	R_{PION} = \frac{\#\text{triplets} \pm 3^{\circ} \text{from peak}}{(\#\text{triplets} \pm 10^{\circ} - \#\text{triplets} \pm 3^{\circ})}
	\end{eqnarray}

	By looking at MC compared to data, the cut criterion was determined to be 0.58 for all SK run periods.  

\subsection{Cherenkov angle}

\begin{figure}[!t]
\centering
\includegraphics[width=2.6in]{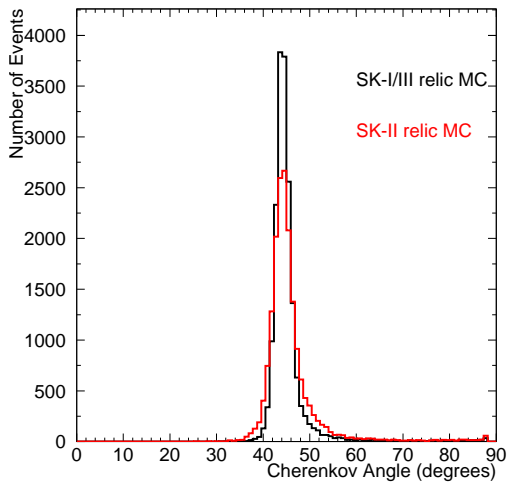}
\caption{Expected Cherenkov angle distribution of SRN events.  This is expected to be model independent; shown is Ando et al.'s LMA \cite{ando2003} relic MC. }
\label{relcangle}
\end{figure}

	Charged particles traveling in pure water at a velocity close to the speed of light in vacuum emit Cherenkov light in a cone shape with an opening angle of about 42 degrees.  Heavier particles, such as muons, pions, and protons, may have a velocity that is smaller if their energies are relatively small, producing Cherenkov light with an opening angle that is less than 42 degrees.  Fig. \ref{relcangle} shows the Cherenkov angle distribution of SN relic MC (all models are very similar).  The Cherenkov angle used is the peak of the distribution of three-hit combination opening angles.  The signal region is determined to be between 38 and 50 degrees.  Events at other Cherenkov angle values are not discarded, but instead they are kept to help extrapolate backgrounds into the signal region, as discussed in the following section.   

\subsection{Other cuts}

	Occasionally events traveling through the OD do not deposit enough light to cause an OD trigger, but still cause a few OD PMT hits.  OD hit clusters are compared in time and space to ID hits and the reconstructed event vertex, and if sufficient correlation exists, the event is rejected.  
	
	The multi-ring cut uses ring counting software developed for atmospheric neutrino and proton decay analyses \cite{sk1oscpar}.  If the software finds that more than one Cherenkov ring exists in the event, the angle between the two rings is calculated, as sometimes the fuzziness of low energy electron Cherenkov rings can cause the event to be mistakenly identified as having more than one ring.  False 2-ring results reconstruct as having a small relative angle.  A 2-ring result separated by more then 60 degrees is judged to be a real multi-ring event, and such events are rejected.  

\begin{table}
\caption[Cut Signal Efficiency (Systematic Error) Summary]{Cut Signal Efficiency (Systematic Error)}
\label{cutsum}
\begin{center}
\begin{tabular}{|c|c|c|c|}
\hline
\textbf{Cut} & \textbf{SK-I} & \textbf{SK-II} & \textbf{SK-III}\\ 
\hline
Noise Reduction & 99$\%$ (1$\%$) & 99$\%$ (1$\%$) & 99$\%$ (1$\%$) \\
Spall + Solar & 88$\%$ (1$\%$) & 87$\%$ (1.4$\%$) & 89$\%$ (1$\%$) \\
Incoming event & 98$\%$ (0.5$\%$) & 95$\%$ (0.3$\%$) & 96$\%$ (0.3$\%$)\\
Pion & 98$\%$ (0.2$\%$) & 97$\%$ (0.5$\%$) & 98$\%$ (0.5$\%$) \\
Cherenkov angle & 95$\%$ (0.4$\%$) & 88$\%$ (3$\%$) & 94$\%$ (0.3$\%$) \\
Other cuts & 98$\%$ (2$\%$) & 98$\%$ (2$\%$) & 98$\%$ (2$\%$) \\
\hline
Total & 78.5$\%$ (2.5$\%$) & 69.2$\%$ (4.0$\%$) & 76.7$\%$ (2.5$\%$) \\
\hline
\end{tabular}
\end{center}
\end{table}

\setcounter{secnumdepth}{3} 

\subsection{Efficiency and systematic error}

	Most cut efficiencies were calculated using relic MC.  Each relic model has slightly different efficiencies; the ones listed in Table \ref{cutsum} are from the LMA model.  The systematic error of most cuts was calculated by comparing LINAC \cite{linacpaper} MC to LINAC data, or by using the values established in the solar analyses when the cuts are similar (noise reduction, incoming event cut).  The OD correlated cut uses a random sample of cosmic ray muon data for OD hit information for determining efficiency.  The efficiency of the spallation cut is a function of position (efficiency is greater near the edges of the detector, where the candidate can be correlated to fewer muons), and the vertex distribution of relic MC is used to estimate the average efficiency.  The error on this estimation is likely dominated by the statistics of our relic MC and spallation random sample, and is rounded up to be conservative.   The efficiency of the solar angle cut is mostly geometric and exact; the only source of systematic error comes from the ratio of events falling in each multiple Coulomb scattering goodness bin, as determined from solar MC.  Values of the systematic errors, along with the cut efficiencies, can be seen in Table \ref{cutsum}.  
	
\begin{figure}[!t]
\centering
\includegraphics[width=2.8in]{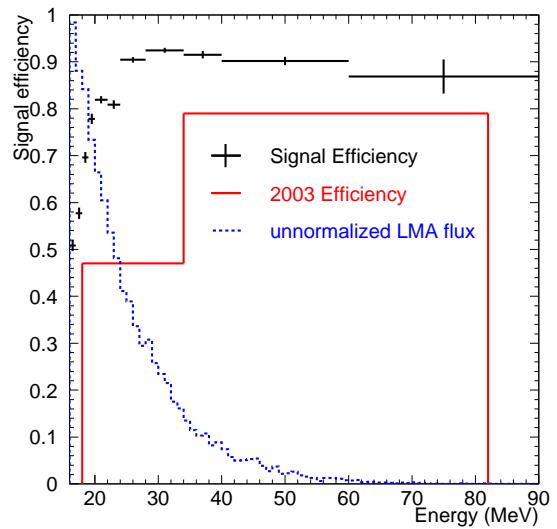}
\caption{Signal efficiency of new analysis compared to 2003 study.  An unnormalized LMA spectrum is overplotted to show the importance of the improvements. }
\label{eff}
\end{figure}

\section{Remaining Backgrounds}

\subsection{Background Sources}

\setcounter{secnumdepth}{1}

	Even after all cuts, some backgrounds remain that must be modeled.  The remaining background can be categorized into four main groupings.  Atmospheric $\nu_{\mu}$ and $\nu_{e}$ charged current (CC) backgrounds were considered in the 2003 SK published SRN paper, and now additional atmospheric neutrino backgrounds are also considered.  Although remaining backgrounds other than the four listed below were found to exist (for example, from multiple and neutral pion production), their contributions were small and their spectrum shapes well described by linear combinations of our four modeled backgrounds.  Thus, any contribution from these sources will be absorbed into the four backgrounds below during the fit.
	
\subsection{1: Atmospheric $\nu_{\mu}$ CC events}

	This is the largest remaining background in our sample.  Atmospheric $\nu_{\mu}$'s and $\bar{\nu}_{\mu}$'s interact in the water of the detector and create a muon via a charged current reaction.  This muon is very low energy, often below Cherenkov threshold, in which case its decay electron cannot be removed like other decay electrons (by correlation to the preceding muon), since the muon is invisible.  This background's energy spectrum is the well known Michel spectrum, slightly modified by resolution effects, and is quite different from the SN relic spectrum.
	
\subsection{2: $\nu_{e}$ CC events}
	
	This background also originates from atmospheric cosmic ray interactions.  Atmospheric $\nu_{e}$'s and $\bar{\nu}_e$'s are indistinguishable from SN relic $\bar{\nu}_e$'s on an individual basis.  Their spectrum is quite different, however.
	
\subsection{3: Atmospheric $\nu$ neutral current (NC) elastic events}

	NC elastic events have an energy spectrum that rises sharply at our lower energy bound, similar to SN relics.  Most are removed by Cherenkov angle reconstruction, but some still leak into our final sample and must be modeled.  With the lowering of the energy threshold from 18 MeV to 16 MeV, this background has become much more relevant.
	
\subsection{4: $\mu/\pi$ production from atmospheric $\nu$}

	This last category is a grouping of two different things, both heavier particles.  First, NC reactions produce charged pions $>$ 200 MeV/C, some of which survive the pion cut.  These remaining events must be modeled.  Included with them, since the spectrum and Cherenkov angle distribution are relatively similar, are surviving muons above Cherenkov threshold.  

\setcounter{secnumdepth}{3}

\subsection{Background Modeling}

	All our backgrounds are modeled using the SK MC, based on GEANT 3 and NEUT \cite{neut}.  Hadronic interactions are simulated using a combination of CALOR and custom SK code.  For the Michel spectrum (used for the $\nu_{\mu}$ CC), we measure the spectrum of decay electrons from cosmic ray muons.  The spectra of the other three remaining backgrounds are determined from the MC, as is the Cherenkov angle distribution of all four backgrounds.  The SN relic signal itself is modeled separately, with its own MC.  Many different models are considered, and different results are calculated for each model.
	
\begin{figure}[!t]
\centering
\includegraphics[width=2.8in]{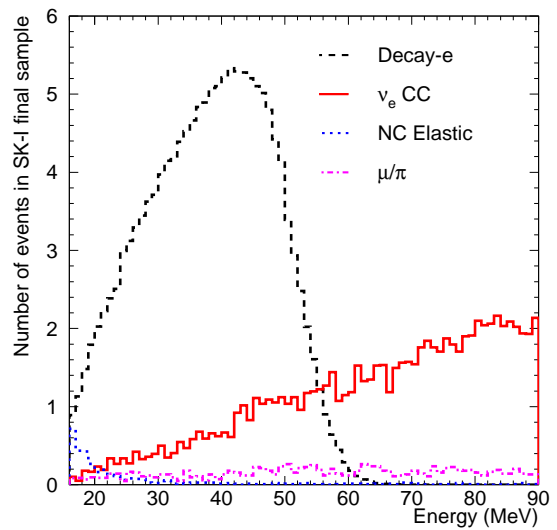}
\caption{Spectra of the four remaining backgrounds in the signal Cherenkov angle region with all reduction cuts applied.  The $\nu_{\mu}$ CC channel is from decay electron data; the other three are from MC.  All are scaled to the SK-I LMA best fit result. }
\label{backs}
\end{figure}

\begin{figure}[!t]
\centering
\includegraphics[width=3in]{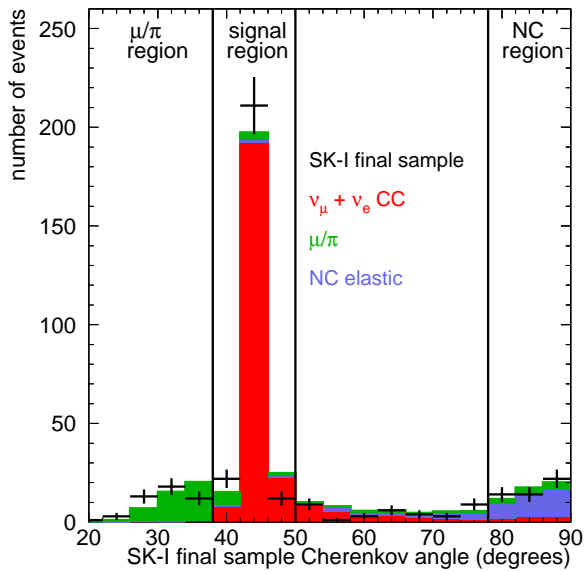}
\caption{Cherenkov angle of SK-I combined final data (all cuts except Cherenkov angle cut applied) overlaid with distributions of the four remaining backgrounds from SK-I MC (same cuts applied).  The division of Cherenkov angle regions is also pictured.}
\label{cangle}
\end{figure}

	Fig. \ref{backs} shows the energy spectrum of the four remaining backgrounds taken from the MC.  Fig. \ref{cangle} shows the Cherenkov angle of the four backgrounds, with no Cherenkov angle cut applied (from MC); it can be seen that the MC has a shape quite similar to the data.  Also, while the CC backgrounds have a Cherenkov angle distribution similar to that of SN relics (38-50 degrees), the NC elastic background mostly reconstructs at high angles (where the Cherenkov angle reconstruction algorithm, which assumes a single ring, misreconstructs events with multiple gammas and roughly isotropic distributions of light), and the $\mu/\pi$ leakage mostly reconstructs at lower angles (as expected from heavier, low energy particles).  Because of these distributions, we have split the data into three regions: the signal region (38-50 degrees); and the two background regions, or `sidebands', consisting of the low (20-38 degrees) $\mu/\pi$ region, and high (78-90 degrees) NC elastic region.  The sidebands are used to normalize the NC elastic and $\mu/\pi$ backgrounds in the signal region, where the SN relic signal would occur.

\section{Limit Extraction}

\subsection{Likelihood fit}

	The relic best fit and upper flux limit are determined by performing an unbinned maximum likelihood fit.  A simultaneous fit is done in all three Cherenkov angle regions.  For each of the Cherenkov angle regions the spectrum of each of the five parameters (SN relic + 4 backgrounds) is parameterized into an analytical function, which is used as the PDF for that parameter.  Each possible reasonable combination of parameters is checked, and the best fit is the combination that maximizes the likelihood, which is determined by the function: 
	
	\begin{eqnarray}	
	\mathcal{L} &=& \prod_{i=1}^{N_{events}} e^{-\sum_{j=1}^{5}{c_j}} \times \sum_{j=1}^{5}{c_{j}pdf_{j}(E_i)} 
	\end{eqnarray}
	
	where $c$ is the coefficient for parameter $j$ representing the number of events for that channel present in all three Cherenkov angle regions combined, which modifies the PDF for that channel, $pdf(E)$.  The likelihood in each Cherenkov angle region is separately calculated, but maximized in conjunction.  The likelihood is maximized for each SK phase separately.

\begin{figure}[!t]
\centering
\includegraphics[width=2.6in]{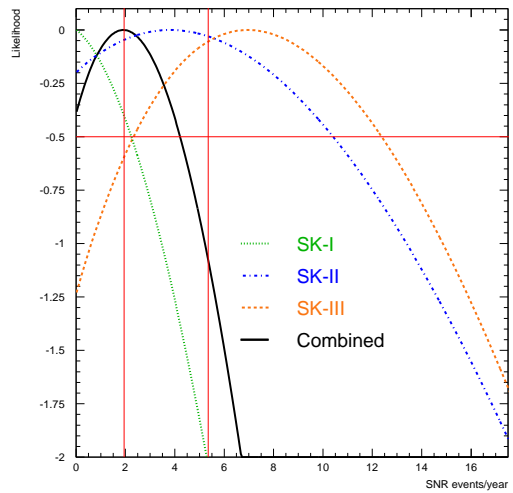}
\caption{Example likelihood curves from the LMA model.  Vertical lines represent best fit and 90 $\%$ C.L. results for the combined fit.}
\label{likes}
\end{figure}

	The 90$\%$ C.L. flux limit is extracted from the likelihood curve (maximum likelihood as a function of number of relic events).  The likelihood curves of SK-I, II, and III are first multiplied together (see Fig. \ref{likes}), then the 90$\%$ C.L. point limit$_{90}$ is determined by the following simple relationship, where \textbf{r} represents the number of SN relic events:
	
	\begin{equation}
	\int_0^{\mbox{limit}_{90}} \! \mathcal{L}(r) \, \mathrm{d}r / \int_0^{\infty} \! \mathcal{L}(r) \, \mathrm{d}r = 0.9
	\end{equation}

\subsection{Systematic error}

	Four systematic errors are considered:
	
\subsubsection{Energy Scale and Resolution Uncertainty}

	The energy scale and energy resolution systematic uncertainties are considered to be uncorrelated.  They are separately calculated, then added in quadrature.  The amount of the energy scale and energy resolution systematic errors are based on the SK solar analyses \cite{sk1solar, sk2solar}, then increased due to the broader energy range and higher energies, and to cover slight uncertainties in the MC.  The energy resolution systematic uncertainty used in our study is 3$\%$, while the energy scale uncertainties used are 2$\%$ for SK-I and SK-III, and 3$\%$ for SK-II.  Inclusion of this systematic error was not found to have an appreciable effect on the results.
	
	To take the energy scale systematic uncertainty into account, the PDFs for four of the five channels (the $\nu_{\mu}$ CC was taken from decay electron data, and is thus not MC, and not susceptible to this error) were shifted by the amount of the uncertainty.  To take into account the energy resolution systematic uncertainty, the energy resolution function from reference \cite{sk2solar} was applied to the likelihood functions to distort them. 
	
	Finally, the full uncertainty was incorporated into the likelihood calculation algorithm as follows.  Let $\epsilon$ be a variable representing the amount of spectral distortion, such that an $\epsilon$ of 1 represents a 1$\sigma$ deviation in the energy scale and resolution; and let $\mathcal{L}(\epsilon)$ be the likelihood evaluated with the likelihoods distorted by an amount $\epsilon$.  Then the new final likelihood $\mathcal{L}'$ is:
	
	\begin{equation}	
	\mathcal{L}' = \frac{1}{\sqrt{2\pi}}\int_{-\infty}^\infty \! e^{-\epsilon^{2}/2} \mathcal{L}(\epsilon) \, \mathrm{d}\epsilon
	\end{equation}
	
	such that the originally calculated likelihood $\mathcal{L}$ is now simply the $\epsilon$ = 0 case.

\subsubsection{Atmospheric $\nu$ Background Systematic Errors}	

	Three of the four remaining backgrounds are modeled by SK MC (the exception being the $\nu_{\mu}$ CC background, which is modeled by decay electron data).  The SK MC is well verified above 100 MeV; however, in the lower energy region of 16-100 MeV that is relevant to our study, the MC accuracy is less well studied.  Little data exists to compare to the MC, especially for the NC elastic mode.  To incorporate these concerns into the analysis, two systematic errors on the $\nu_{e}$ CC and NC elastic backgrounds were included.  These systematics were designed to cover any remaining uncertainties in the MC.  As for the other two background channels, the $\nu_{\mu}$ CC comes from real data and therefore is not vulnerable to MC inaccuracies, and the $\mu/\pi$ channel is by far the least relevant of the four channels, and therefore the safest to neglect.  

	For the NC elastic channel, the greatest potential for error lies in the relative normalization across Cherenkov angle regions.  The amount of NC elastic found in the 78-90 degree region determines the amount of the fit in the signal region.  If the ratio of the amount of NC elastic in the high angle region compared to the signal region as determined by MC is off by some factor, this will tend to have a larger influence on the result than a distortion of the spectrum of the NC elastic in the signal region by the same factor.  

	For SK-I, the NC elastic MC has 7.4$\%$ of events occur in the signal region, while 87$\%$ of events occur in the high angle region.  Considered is a 100$\%$ change in the normalization of the number of NC elastic events in the signal region, with all changes to be correspondingly balanced by a change in the normalization of the high angle region.  Thus, a +1$\sigma$ effect would be for 14.8$\%$ of events to occur in the signal region, and 79.6$\%$ of events to occur in the high angle region, while a -1$\sigma$ effect would be for no NC elastic events to occur in the signal region, and 94.4$\%$ of the events to occur in the high angle region.  As we can't have less than 0 events in the signal region, there is a physical boundary at -1$\sigma$.  Thus, we applied the error asymmetrically from -1 to +3 $\sigma$.

	For the $\nu_{e}$ CC channel, the relative normalization across Cherenkov angle regions is not a concern, as more than 99$\%$ of the events fall into the signal region, so instead a distortion of the spectrum in the signal region was considered.  The $\nu_{e}$ CC PDF in the signal region was distorted as follows:

	\begin{eqnarray}	
	\mbox{PDF}_{new} = \mbox{PDF}_{old} \times N(1+\frac{0.5{\epsilon}(E-16)}{74 \mbox{ MeV}}) 
	\end{eqnarray}

where E is the energy (MeV), $\epsilon$ is the magnitude of the spectral distortion, and N is a constant such that the new PDF is normalized the same as the old.

	A -2$\sigma$ effect would reduce the PDF to 0 at 90 MeV.  This was considered to be unphysical, and thus a -1$\sigma$ effect was considered the physical lower bound, forcing us to apply this error in the same asymmetric manner as the NC elastic, from -1 to +3$\sigma$.
	
	We applied the background channel systematics using an integral method with a weighting function (Bayesian prior).  However, as our error is applied over an asymmetric region (-1 to +3 $\sigma$), we could not use a symmetric Gaussian (such as we use in the next section) without introducing a bias in our result.  Instead, we used a weighted Gaussian that as closely as possible maintained the properties of the symmetric Gaussian.
	
	Specifically, in the case of a symmetric error (i.e., -$\infty$ to $\infty$), a normal Gaussian would have the following properties: (1) expectation value (first moment) = 0, (2) variance (second moment) = $\sigma^2$.  Our weighted Gaussian was constructed such that its properties (integrated over $\sigma$ from -1 to 3) were the same as the symmetric case.  This weighted Gaussian was normalized to 1 and used in both the $\nu_{e}$ CC and NC elastic cases, as their integration ranges were identical.
	
	The integral method of applying the systematic error is as follows.  Let G($\sigma$) be our weighted Gaussian function described above.  Let $\mathcal{L}(r,\sigma)$ be the likelihood as a function of SN relic events and whichever systematic error is under consideration.  Then, the likelihood after the application of the systematic error of either of the background channels is:
	
	\begin{eqnarray}	
	\mathcal{L}(r)_{new}= \int_{-1}^3 \mathcal{L}(r,\sigma)_{old} G(\sigma)\mathrm{d}\sigma
	\end{eqnarray}

	Inclusion of the atmospheric $\nu$ background systematics changed the result (towards a less stringent limit) by less than 6$\%$.

\subsubsection{Energy Independent Efficiency Systematic Error}

	The energy independent portion of the efficiency sytematic error is assumed uncorrelated.  The systematic error of each cut, as shown in Table \ref{cutsum}, is added in quadrature, to get the total efficiency systematic error of the reduction.  The total efficiency systematic error also includes the uncertainty in the fiducial volume cut, inverse beta cross section, and livetime calculation, as shown in Table \ref{syssum}.
	
\begin{table}
\caption[Total Efficiency Systematic Error]{Total Efficiency Systematic Error}
\label{syssum}
\begin{center}
\begin{tabular}{|c|c|c|c|}
\hline
\textbf{Error Source} & \textbf{SK-I} & \textbf{SK-II} & \textbf{SK-III}\\ 
\hline
Cut reduction & 3.1$\%$ & 4.4$\%$ & 3.1$\%$ \\
Fiducial Volume & 1.3$\%$  & 1.1$\%$ & 1.0$\%$ \\
Cross Section & 1.0$\%$ & 1.0$\%$ & 1.0$\%$\\
Livetime & 0.1$\%$ & 0.1$\%$ & 0.1$\%$ \\
\hline
Total & 3.51$\%$  & 4.65$\%$  & 3.41$\%$  \\
\hline
\end{tabular}
\end{center}
\end{table}

	The fiducial volume and livetime calculation systematic errors are the same as, and taken from, the SK solar analyses \cite{sk1solar,sk2solar}.
	
	To apply this systematic error, the likelihood curve (likelihood as a function of number of SN relic events) is modified.  The modification for each value of the likelihood is as follows, where $\epsilon$ is the efficiency, $\sigma$ is the efficiency systematic error, P($\epsilon$) is a probability function in the shape of a Gaussian centered on $\epsilon_{0}$ with width $\epsilon_{0}\sigma$, r is the number of relic events seen in the data, and R is the number of relic events actually interacting in the detector, such that r = $\epsilon$R:

	\begin{equation}	
	\mathcal{L}'(R) = \int_{\epsilon = 0}^1 \! \mathcal{L}({\epsilon}R)P(\epsilon){\epsilon} \, \mathrm{d}\epsilon
	\end{equation}

	This error is applied last, after all other calculations are complete.  Inclusion of this systematic error changed the result (making the limit less stringent) by a few percent.

\section{The result}

\begin{figure}[!t]
\centering
\includegraphics[width=3in]{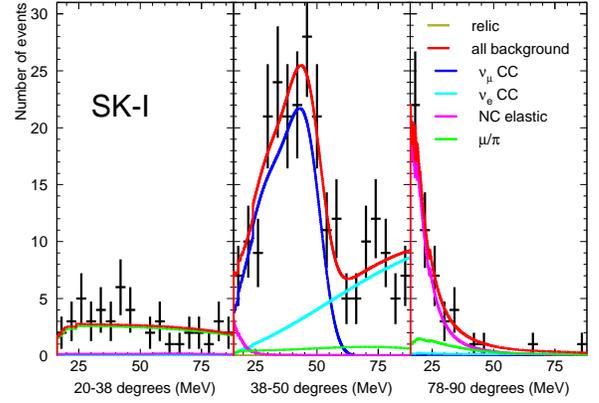}
\caption{SK-I LMA best fit result.  The relic best fit is negative, so a relic fit of 0 is shown.}
\label{sk1fit}
\end{figure}

\begin{figure}[!t]
\centering
\includegraphics[width=3in]{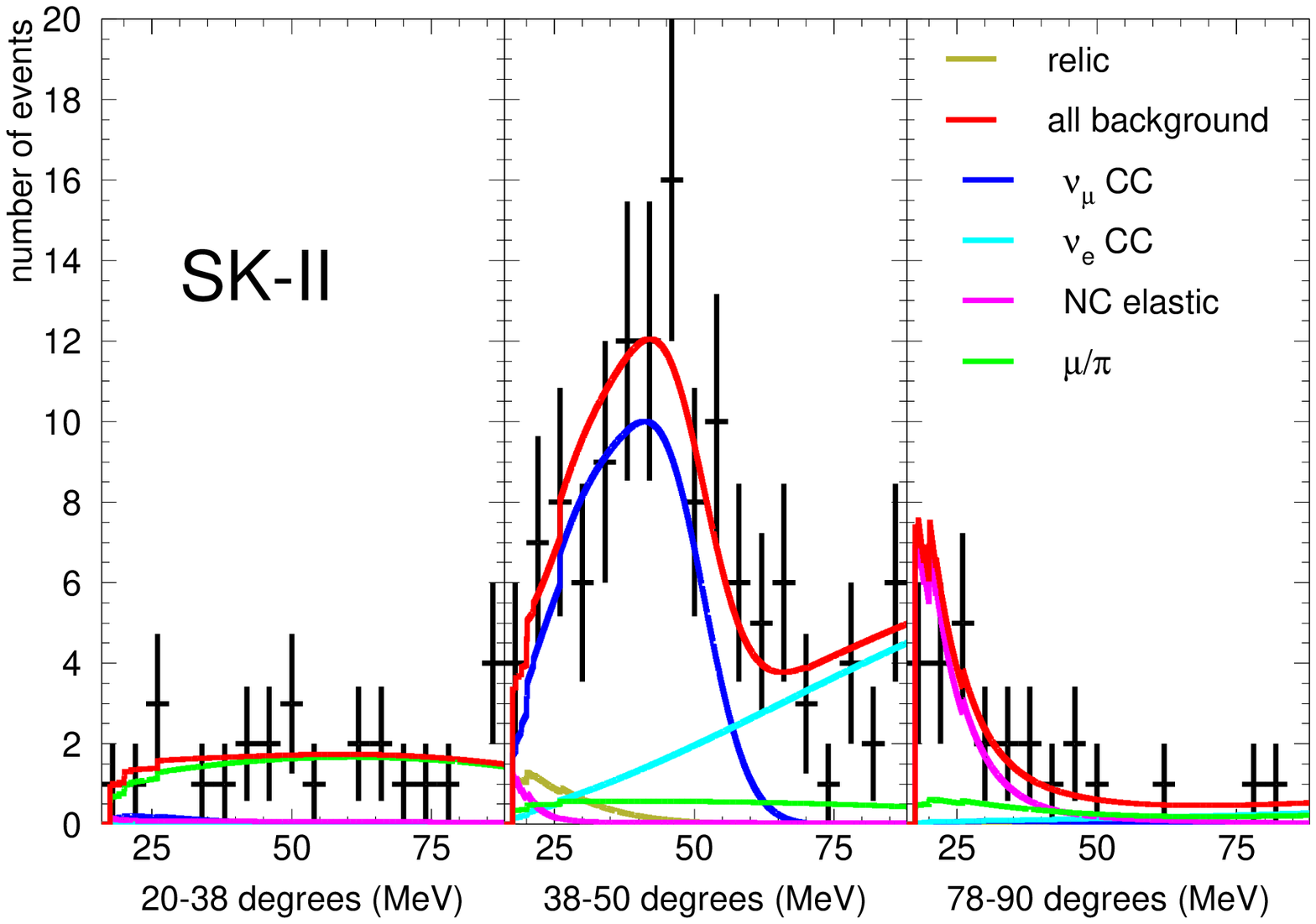}
\caption{SK-II LMA best fit result.  The relic best fit is 3.05 events per year interacting in the detector (before reduction efficiencies).}
\label{sk2fit}
\end{figure}

\begin{figure}[!t]
\centering
\includegraphics[width=3in]{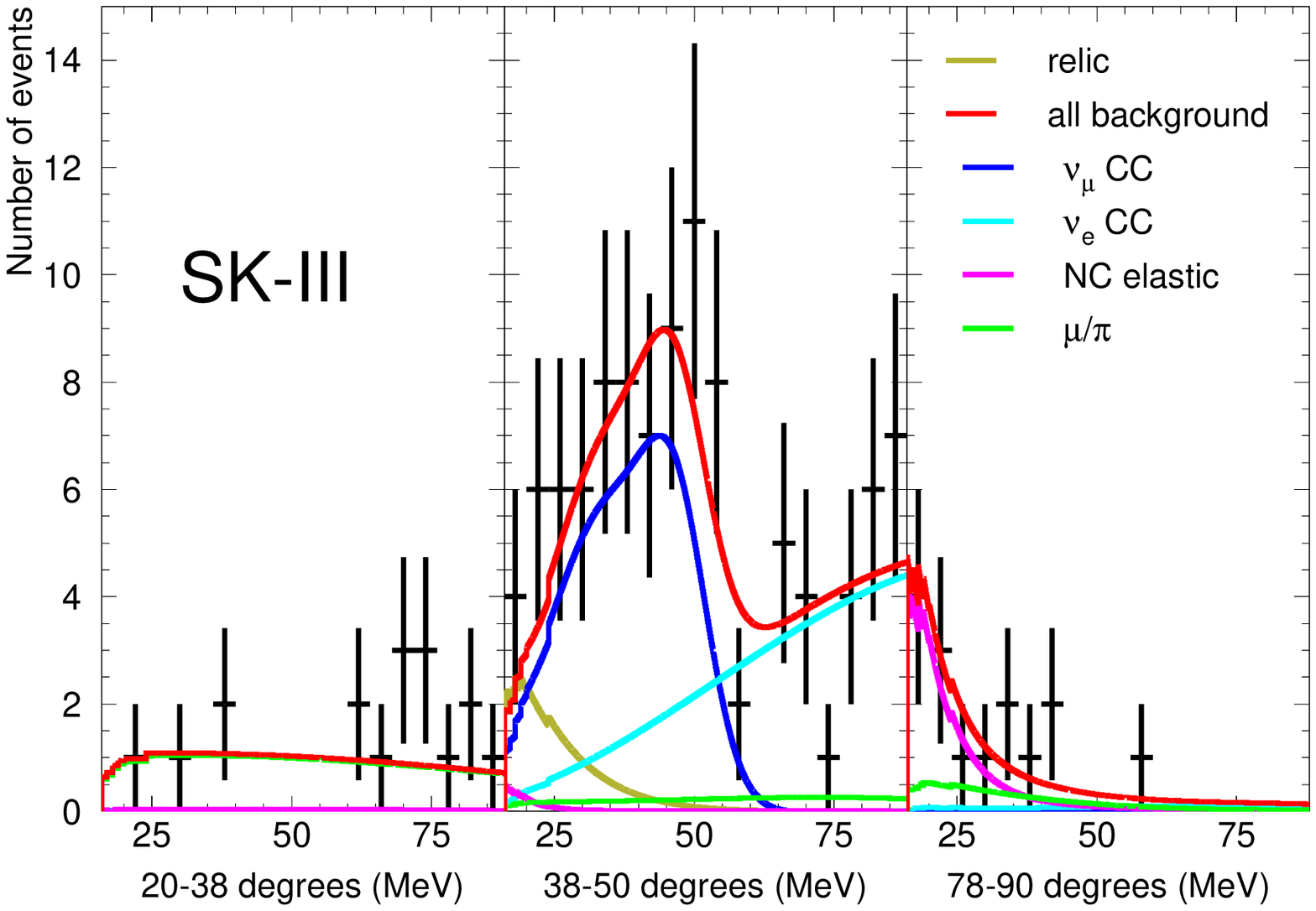}
\caption{SK-III LMA best fit result.  The relic best fit is 6.9 events per year interacting in the detector (before reduction efficiencies).}
\label{sk3fit}
\end{figure}

	Figures \ref{sk1fit}, \ref{sk2fit}, and \ref{sk3fit} show the final sample data, overlaid with the best fit results for the LMA model.  For SK-I, the relic best fit is negative and unphysical (for all models), so zero relic contribution is shown.  However, for all models SK-II and SK-III find a positive indication (not significant) of a relic component.  The 90$\%$ C.L. upper flux limits differ for each model, and are summarized in Table \ref{results}.  

\begin{table}
\caption[90$\%$ C.L. flux limit ($\bar{\nu}_e$ cm$^{-2}$ s$^{-1}$), $E_{\nu} >$ 17.3 MeV]{90 $\%$ C.L. flux limit ($\bar{\nu}$ cm$^{-2}$ s$^{-1}$), $E_{\nu} >$ 17.3 MeV}
\label{results}
\begin{center}
\begin{tabular}{|c|c|c|c|c|c|}
\hline
\textbf{Model} & \textbf{SK-I} & \textbf{SK-II} & \textbf{SK-III} & \textbf{All} & \textbf{Predicted}\\ 
\hline
Gas Infall (97) & $<$2.1 & $<$7.5 & $<$7.8 & $<$2.8 & 0.3\\
Chemical (97) & $<$2.2 & $<$7.2 & $<$7.8 & $<$2.8 & 0.6\\
Heavy Metal (00) & $<$2.2 & $<$7.4 & $<$7.8 & $<$2.8 & $<$ 1.8\\
LMA (03) & $<$2.5 & $<$7.7 & $<$8.0 & $<$2.9 & 1.7\\
Failed SN (09) & $<$2.4 & $<$8.0 & $<$8.4 & $<$3.0 & 0.7\\
6 MeV (09) & $<$2.7 & $<$7.4 & $<$8.7 & $<$3.1 & 1.5\\ 
\hline
\end{tabular}
\end{center}
\end{table}

\subsection{Comparison to previous study}

	The new analysis has a lower energy threshold than the 2003 study, so the results are not immediately comparable.  Furthermore, the previous study presented its result in a model independent fashion due to the similarity of the results.  Although we verified that with an 18 MeV energy threshold the SK-I results were similar enough to justify a model-independent approach, the lower energy threshold and new data required model dependent results.  For comparison purposes, we consider the LMA model.  The limit is 2.9 $\bar{\nu}$ cm$^{-2}$ s$^{-1}$ $>$ 16 MeV (positron energy), which is equivalent to 2.0 $\bar{\nu}$ cm$^{-2}$ s$^{-1}$ $>$ 18 MeV positron energy.  It is notable that this result is less stringent than the 2003 result of 1.2 $\bar{\nu}$ cm$^{-2}$ s$^{-1}$ positron energy $>$ 18 MeV.  There are multiple reasons for this.
	
	First, a $0^{th}$ order approximation of the inverse beta cross section was then used.  Now, the full cross section from \cite{strumia} is used.  This raises the limit by about $8\%$.  If events with post-activity are also removed, the old-style analysis limit becomes 1.35 cm$^{-2}$ s$^{-1}$.  Furthermore, the binned $\chi^2$ method used assumed Gaussian statistics, while Poissonian statistics are more appropriate considering the low statistics.  This alone would change the limit from 1.2 to 1.7 cm$^{-2}$ s$^{-1}$.  When all these corrections are combined, the original analysis result of 1.2 $\bar{\nu}$ cm$^{-2}$ s$^{-1}$ instead becomes 1.9 $\bar{\nu}$ cm$^{-2}$ s$^{-1}$.  
	
	With our improved analysis, if we neglect atmospheric $\nu$ background systematics (which were not fully included in the 2003 study), the SK-I only LMA result is 1.6 $\bar{\nu}$ cm$^{-2}$ s$^{-1}$ ($>$ 18 MeV positron energy), which is more stringent than the published analysis with these corrections.  However, the SK-II and SK-III data shows a hint of a signal, which causes the limit to become less stringent when all the data are combined, for the final LMA result (with all systematics) of 2.0 $\bar{\nu}$ cm$^{-2}$ s$^{-1}$ $>$ 18 MeV positron energy, or 2.9 $\bar{\nu}$ cm$^{-2}$ s$^{-1}$ $>$ 16 MeV positron energy.  

\subsection{Typical SN $\nu$ emission limit}

\begin{figure}[!t]
\centering
\includegraphics[width=3in]{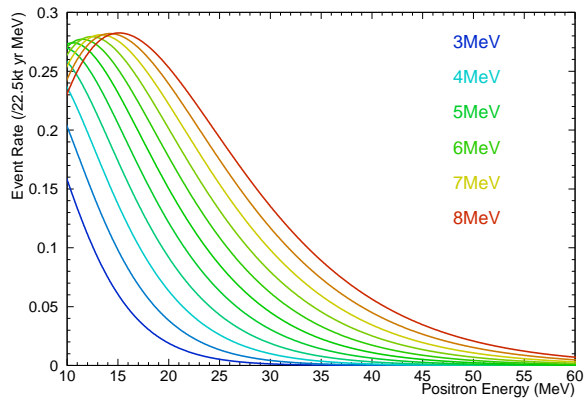}
\caption{True positron spectra in SK for each neutrino temperature, from 3 MeV to 8 MeV in 0.5 MeV steps (SN $\bar{\nu}_e$ luminosity of 5$\times 10^{52}$ ergs assumed).}
\label{espec}
\end{figure}

\begin{figure}[!t]
\includegraphics[width=3.6in]{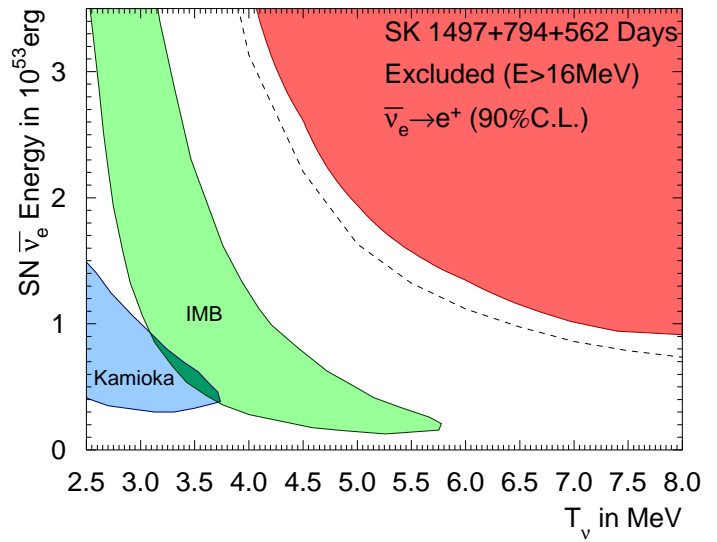}
\caption{Results plotted as an exclusion contour in SN neutrino luminosity vs. neutrino temperature parameter space.  The IMB and Kamiokande allowed areas for 1987A data are shown (originally from \cite{raffelt}) along with our new 90$\%$ C.L. result.  The dashed line shows the individual 90$\%$ C.L. results of each temperature considered separately, which is not a true 2-D exclusion contour.  Results are in the form of Fig. 6 from \cite{Beacom:2010kk}.}
\label{contour}
\end{figure}

\begin{figure}[!t]
\centering
\includegraphics[width=3.6in]{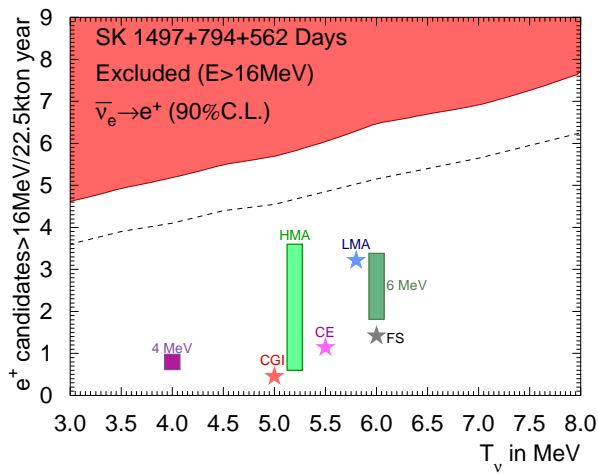}
\caption{Exclusion contour plotted in a parameter space of SRN event rate vs. neutrino temperature.  The red contour shows our 90$\%$ C.L. result.  The dashed line shows the individual 90$\%$ C.L. results of each temperature considered separately, which is not a true 2-D exclusion contour.  CGI is Cosmic Gas Infall model, HMA is Heavy Metal Abundance model, CE is Chemical Evolution model, LMA is Large Mixing Angle model, FS is Failed Supernova model, and the 6 and 4 MeV cases are from \cite{isdetect}.  For the 4 and 6 MeV cases a total uncertainty is provided and shown, and the HMA model gives a range which is shown.  Other models have no given range or uncertainty and are represented by a star.}
\label{contour2}
\end{figure}

	Most of the elements involved in a comprehensive prediction of the SRN flux are now fairly well known \cite{Beacom:2010kk} (e.g. initial mass functions, cosmic star formation history, Hubble expansion, etc.), and thus we can parameterize typical supernova neutrino emission using two effective parameters \cite{yukselando,isdetect}: $\bar{\nu}_e$ luminosity from a typical supernova, and average emitted $\bar{\nu}_e$ energy.
	
	To calculate this result, SRN MC samples were created with a Fermi-Dirac (FD) emission spectrum at multiple temperatures (see Fig. \ref{espec}).  The Ando et al. model \cite{ando2003} corresponds to a FD spectrum of close to 6 MeV and a $\bar{\nu}_e$ luminosity of 7$\times 10^{45}$ J.  MC was created from 2.5 to 8 MeV in 0.5 MeV steps.  PDFs were made of these spectra, and the final data sample was fit, as previously described, with these new likelihoods in the place of the other SRN models.  The following constants were assumed: $\Omega_{\mbox{\tiny{m}}} = 0.3$, $\Omega_{\Lambda} = 0.7$, and c/$\mbox{H}_0$ = 4228 Mpc.  The result is independent of the initial mass function used.

	 Our result is shown in Fig. \ref{contour} and Fig. \ref{contour2}.  Also plotted for convenience is the 1-D line representing the results of each neutrino temperature analysed separately (the numerical values are given in Table \ref{2dsum}).

\begin{table}
\caption[Summary of limit values.]{Summary of limit values.}
\label{2dsum}
\begin{center}
\begin{tabular}{|c|c|}
\hline
\textbf{$\bar{\nu}_e$ Temperature} (MeV) & \textbf{Energy} ($\times 10^{53}$ ergs) \\
\hline
3.0 & 9.0 \\
3.5 & 4.8 \\
4.0 & 2.9 \\
4.5 & 2.1 \\
5.0 & 1.5 \\
5.5 & 1.3 \\
6.0 & 1.1 \\
6.5 & 0.92 \\
7.0 & 0.82 \\
7.5 & 0.75 \\ 
8.0 & 0.70 \\
\hline
\end{tabular}
\end{center}
\end{table}

\section{Conclusion and Future}

	In summary, we have made large improvements to the 2003 SK relic analysis.  The cuts have been optimized, greatly increasing efficiency, up to around $\sim75\%$, and lowering the energy threshold down to 16 MeV.  New remaining backgrounds are now considered and modeled, and we have almost twice as much data in our final sample.  A new maximum likelihood fit with multiple Cherenkov angle regions is utilized to extract a set of model-dependent upper flux limits.  Further, we have expressed our results in a model-independent fashion using two effective parameters to model typical SN neutrino emission.

	Although our limit is tantalizingly close to the best theoretical predictions, no signal has so far been detected.  While SK is expected to continue data taking for many more years, future sensitivity improvements will be slow, since SK-I/II/III's exposure is already 176 kt-years, and the analysis is now highly optimized.  50 kt-years of SK-IV data already exists, and the new SK-IV electronics structure may allow for some further background reduction, as decay electrons from atmospheric $\nu_{\mu}$ CC interactions can now be tagged by detection of prompt gamma emissions with higher efficiency.   Further improvement would be possible with new methods, such as the doping of SK water with gadolinium, which could lower the energy threshold and backgrounds dramatically and could allow detection of the SRN signal within five years \cite{gadzooks}.  

\section{Acknowledgements}

	We gratefully acknowledge the cooperation of the Kamioka Mining and Smelting Company. The Super-Kamiokande experiment has been built and operated from funding by the Japanese Ministry of Education, Culture, Sports, Science and Technology, the United States Department of Energy, and the U.S. National Science Foundation. Some of us have been supported by funds from the Korean Research Foundation (BK21), the National Research Foundation of Korea(NRF-20110024009), the Japan Society for the Promotion of Science, and the National Natural Science Foundation of China (grants 10875062 and 10911140109).

\bibliography{skrelic}

\begin{thebibliography}{34}%
\makeatletter
\providecommand \@ifxundefined [1]{%
 \@ifx{#1\undefined}
}%
\providecommand \@ifnum [1]{%
 \ifnum #1\expandafter \@firstoftwo
 \else \expandafter \@secondoftwo
 \fi
}%
\providecommand \@ifx [1]{%
 \ifx #1\expandafter \@firstoftwo
 \else \expandafter \@secondoftwo
 \fi
}%
\providecommand \natexlab [1]{#1}%
\providecommand \enquote  [1]{``#1''}%
\providecommand \bibnamefont  [1]{#1}%
\providecommand \bibfnamefont [1]{#1}%
\providecommand \citenamefont [1]{#1}%
\providecommand \href@noop [0]{\@secondoftwo}%
\providecommand \href [0]{\begingroup \@sanitize@url \@href}%
\providecommand \@href[1]{\@@startlink{#1}\@@href}%
\providecommand \@@href[1]{\endgroup#1\@@endlink}%
\providecommand \@sanitize@url [0]{\catcode `\\12\catcode `\$12\catcode
  `\&12\catcode `\#12\catcode `\^12\catcode `\_12\catcode `\%12\relax}%
\providecommand \@@startlink[1]{}%
\providecommand \@@endlink[0]{}%
\providecommand \url  [0]{\begingroup\@sanitize@url \@url }%
\providecommand \@url [1]{\endgroup\@href {#1}{\urlprefix }}%
\providecommand \urlprefix  [0]{URL }%
\providecommand \Eprint [0]{\href }%
\providecommand \doibase [0]{http://dx.doi.org/}%
\providecommand \selectlanguage [0]{\@gobble}%
\providecommand \bibinfo  [0]{\@secondoftwo}%
\providecommand \bibfield  [0]{\@secondoftwo}%
\providecommand \translation [1]{[#1]}%
\providecommand \BibitemOpen [0]{}%
\providecommand \bibitemStop [0]{}%
\providecommand \bibitemNoStop [0]{.\EOS\space}%
\providecommand \EOS [0]{\spacefactor3000\relax}%
\providecommand \BibitemShut  [1]{\csname bibitem#1\endcsname}%
\let\auto@bib@innerbib\@empty
\bibitem [{\citenamefont {{E. Cappellaro, R. Evans, and M.
  Turatto}}(1999)}]{snrate}%
  \BibitemOpen
  \bibfield  {author} {\bibinfo {author} {\bibnamefont {{E. Cappellaro, R.
  Evans, and M. Turatto}}},\ }\href@noop {} {\bibfield  {journal} {\bibinfo
  {journal} {Astron. Astrophys.}\ }\textbf {\bibinfo {volume} {351}},\ \bibinfo
  {pages} {459} (\bibinfo {year} {1999})}\BibitemShut {NoStop}%
\bibitem [{\citenamefont {{G.S. Bisnovatyi-kogan and Z.F.
  Seidov}}(1984)}]{bisno}%
  \BibitemOpen
  \bibfield  {author} {\bibinfo {author} {\bibnamefont {{G.S. Bisnovatyi-kogan
  and Z.F. Seidov}}},\ }\href@noop {} {\bibfield  {journal} {\bibinfo
  {journal} {Ann. N.Y. Acad. Sci.}\ }\textbf {\bibinfo {volume} {422}},\
  \bibinfo {pages} {319} (\bibinfo {year} {1984})}\BibitemShut {NoStop}%
\bibitem [{\citenamefont {Krauss}\ \emph {et~al.}(1984)\citenamefont {Krauss},
  \citenamefont {Glashow},\ and\ \citenamefont {Schramm}}]{krauss}%
  \BibitemOpen
  \bibfield  {author} {\bibinfo {author} {\bibfnamefont {L.~M.}\ \bibnamefont
  {Krauss}}, \bibinfo {author} {\bibfnamefont {S.~L.}\ \bibnamefont {Glashow}},
  \ and\ \bibinfo {author} {\bibfnamefont {D.~N.}\ \bibnamefont {Schramm}},\
  }\href@noop {} {\bibfield  {journal} {\bibinfo  {journal} {Nature}\ }\textbf
  {\bibinfo {volume} {310}},\ \bibinfo {pages} {191} (\bibinfo {year}
  {1984})}\BibitemShut {NoStop}%
\bibitem [{\citenamefont {{Woosley}}\ \emph {et~al.}(1986)\citenamefont
  {{Woosley}}, \citenamefont {{Wilson}},\ and\ \citenamefont
  {{Mayle}}}]{woosley}%
  \BibitemOpen
  \bibfield  {author} {\bibinfo {author} {\bibfnamefont {S.~E.}\ \bibnamefont
  {{Woosley}}}, \bibinfo {author} {\bibfnamefont {J.~R.}\ \bibnamefont
  {{Wilson}}}, \ and\ \bibinfo {author} {\bibfnamefont {R.}~\bibnamefont
  {{Mayle}}},\ }\href {\doibase 10.1086/163968} {\bibfield  {journal} {\bibinfo
   {journal} {Astrophysical Journal}\ }\textbf {\bibinfo {volume} {302}},\
  \bibinfo {pages} {19} (\bibinfo {year} {1986})}\BibitemShut {NoStop}%
\bibitem [{\citenamefont {{Totani}}\ \emph {et~al.}(1996)\citenamefont
  {{Totani}}, \citenamefont {{Sato}},\ and\ \citenamefont
  {{Yoshii}}}]{totani1996}%
  \BibitemOpen
  \bibfield  {author} {\bibinfo {author} {\bibfnamefont {T.}~\bibnamefont
  {{Totani}}}, \bibinfo {author} {\bibfnamefont {K.}~\bibnamefont {{Sato}}}, \
  and\ \bibinfo {author} {\bibfnamefont {Y.}~\bibnamefont {{Yoshii}}},\ }\href
  {\doibase 10.1086/176970} {\bibfield  {journal} {\bibinfo  {journal}
  {Astrophysical Journal}\ }\textbf {\bibinfo {volume} {460}},\ \bibinfo
  {pages} {303} (\bibinfo {year} {1996})},\ \Eprint
  {http://arxiv.org/abs/arXiv:astro-ph/9509130} {arXiv:astro-ph/9509130}
  \BibitemShut {NoStop}%
\bibitem [{\citenamefont {Malaney}(1997)}]{malaney}%
  \BibitemOpen
  \bibfield  {author} {\bibinfo {author} {\bibfnamefont {R.~A.}\ \bibnamefont
  {Malaney}},\ }\href {\doibase DOI: 10.1016/S0927-6505(97)00012-1} {\bibfield
  {journal} {\bibinfo  {journal} {Astroparticle Physics}\ }\textbf {\bibinfo
  {volume} {7}},\ \bibinfo {pages} {125 } (\bibinfo {year} {1997})}\BibitemShut
  {NoStop}%
\bibitem [{\citenamefont {Hartmann}\ and\ \citenamefont
  {Woosley}(1997)}]{hartmann}%
  \BibitemOpen
  \bibfield  {author} {\bibinfo {author} {\bibfnamefont {D.~H.}\ \bibnamefont
  {Hartmann}}\ and\ \bibinfo {author} {\bibfnamefont {S.~E.}\ \bibnamefont
  {Woosley}},\ }\href {\doibase DOI: 10.1016/S0927-6505(97)00018-2} {\bibfield
  {journal} {\bibinfo  {journal} {Astroparticle Physics}\ }\textbf {\bibinfo
  {volume} {7}},\ \bibinfo {pages} {137 } (\bibinfo {year} {1997})}\BibitemShut
  {NoStop}%
\bibitem [{\citenamefont {Kaplinghat}\ \emph {et~al.}(2000)\citenamefont
  {Kaplinghat}, \citenamefont {Steigman},\ and\ \citenamefont
  {Walker}}]{kap2000}%
  \BibitemOpen
  \bibfield  {author} {\bibinfo {author} {\bibfnamefont {M.}~\bibnamefont
  {Kaplinghat}}, \bibinfo {author} {\bibfnamefont {G.}~\bibnamefont
  {Steigman}}, \ and\ \bibinfo {author} {\bibfnamefont {T.~P.}\ \bibnamefont
  {Walker}},\ }\href {\doibase 10.1103/PhysRevD.62.043001} {\bibfield
  {journal} {\bibinfo  {journal} {Phys. Rev. D}\ }\textbf {\bibinfo {volume}
  {62}},\ \bibinfo {pages} {043001} (\bibinfo {year} {2000})}\BibitemShut
  {NoStop}%
\bibitem [{\citenamefont {Ando}\ \emph {et~al.}(2003)\citenamefont {Ando},
  \citenamefont {Sato},\ and\ \citenamefont {Totani}}]{ando2003}%
  \BibitemOpen
  \bibfield  {author} {\bibinfo {author} {\bibfnamefont {S.}~\bibnamefont
  {Ando}}, \bibinfo {author} {\bibfnamefont {K.}~\bibnamefont {Sato}}, \ and\
  \bibinfo {author} {\bibfnamefont {T.}~\bibnamefont {Totani}},\ }\href
  {\doibase DOI: 10.1016/S0927-6505(02)00152-4} {\bibfield  {journal} {\bibinfo
   {journal} {Astroparticle Physics}\ }\textbf {\bibinfo {volume} {18}},\
  \bibinfo {pages} {307 } (\bibinfo {year} {2003})}\BibitemShut {NoStop}%
\bibitem [{Note1()}]{Note1}%
  \BibitemOpen
  \bibinfo {note} {The flux of the LMA model is increased by a factor of 2.56
  from the paper, a revision introduced at NNN05}\BibitemShut {NoStop}%
\bibitem [{\citenamefont {Lunardini}(2009)}]{lunardini}%
  \BibitemOpen
  \bibfield  {author} {\bibinfo {author} {\bibfnamefont {C.}~\bibnamefont
  {Lunardini}},\ }\href {\doibase 10.1103/PhysRevLett.102.231101} {\bibfield
  {journal} {\bibinfo  {journal} {Phys. Rev. Lett.}\ }\textbf {\bibinfo
  {volume} {102}},\ \bibinfo {pages} {231101} (\bibinfo {year}
  {2009})}\BibitemShut {NoStop}%
\bibitem [{Note2()}]{Note2}%
  \BibitemOpen
  \bibinfo {note} {Assumed parameters are: Failed SN rate = 22$\%$, EoS =
  Lattimer-Swesty, and survival probability = 68$\%$}\BibitemShut {NoStop}%
\bibitem [{\citenamefont {Horiuchi}\ \emph {et~al.}(2009)\citenamefont
  {Horiuchi}, \citenamefont {Beacom},\ and\ \citenamefont {Dwek}}]{isdetect}%
  \BibitemOpen
  \bibfield  {author} {\bibinfo {author} {\bibfnamefont {S.}~\bibnamefont
  {Horiuchi}}, \bibinfo {author} {\bibfnamefont {J.~F.}\ \bibnamefont
  {Beacom}}, \ and\ \bibinfo {author} {\bibfnamefont {E.}~\bibnamefont
  {Dwek}},\ }\href {\doibase 10.1103/PhysRevD.79.083013} {\bibfield  {journal}
  {\bibinfo  {journal} {Phys. Rev. D}\ }\textbf {\bibinfo {volume} {79}},\
  \bibinfo {pages} {083013} (\bibinfo {year} {2009})}\BibitemShut {NoStop}%
\bibitem [{\citenamefont {{M. Malek et al.}}(2003)}]{malekpaper}%
  \BibitemOpen
  \bibfield  {author} {\bibinfo {author} {\bibnamefont {{M. Malek et al.}}}
  (\bibinfo {collaboration} {{The Super-Kamiokande Collaboration}}),\ }\href
  {\doibase 10.1103/PhysRevLett.90.061101} {\bibfield  {journal} {\bibinfo
  {journal} {Phys. Rev. Lett.}\ }\textbf {\bibinfo {volume} {90}},\ \bibinfo
  {pages} {061101} (\bibinfo {year} {2003})},\ \Eprint
  {http://arxiv.org/abs/hep-ex/0209028} {arXiv:hep-ex/0209028} \BibitemShut
  {NoStop}%
\bibitem [{\citenamefont {{W. Zhang et al.}}(1988)}]{srn88}%
  \BibitemOpen
  \bibfield  {author} {\bibinfo {author} {\bibnamefont {{W. Zhang et al.}}}
  (\bibinfo {collaboration} {{The Kamiokande Collaboration}}),\ }\href
  {\doibase 10.1103/PhysRevLett.61.385} {\bibfield  {journal} {\bibinfo
  {journal} {Phys. Rev. Lett.}\ }\textbf {\bibinfo {volume} {61}},\ \bibinfo
  {pages} {385} (\bibinfo {year} {1988})}\BibitemShut {NoStop}%
\bibitem [{\citenamefont {Aharmim}\ \emph {et~al.}(2006)\citenamefont {Aharmim}
  \emph {et~al.}}]{snorelic}%
  \BibitemOpen
  \bibfield  {author} {\bibinfo {author} {\bibfnamefont {B.}~\bibnamefont
  {Aharmim}} \emph {et~al.} (\bibinfo {collaboration} {SNO}),\ }\href {\doibase
  10.1086/508768} {\bibfield  {journal} {\bibinfo  {journal} {Astrophys. J.}\
  }\textbf {\bibinfo {volume} {653}},\ \bibinfo {pages} {1545} (\bibinfo {year}
  {2006})},\ \Eprint {http://arxiv.org/abs/hep-ex/0607010}
  {arXiv:hep-ex/0607010} \BibitemShut {NoStop}%
\bibitem [{\citenamefont {{The KamLAND Collaboration}}(2011)}]{kamdsnb}%
  \BibitemOpen
  \bibfield  {author} {\bibinfo {author} {\bibnamefont {{The KamLAND
  Collaboration}}},\ }\href@noop {} {\  (\bibinfo {year} {2011})},\ \Eprint
  {http://arxiv.org/abs/1105.3516} {arXiv:1105.3516 [astro-ph.HE]} \BibitemShut
  {NoStop}%
\bibitem [{\citenamefont {Strigari}\ \emph {et~al.}(2004)\citenamefont
  {Strigari}, \citenamefont {Kaplinghat}, \citenamefont {Steigman},\ and\
  \citenamefont {Walker}}]{strigkaprelic}%
  \BibitemOpen
  \bibfield  {author} {\bibinfo {author} {\bibfnamefont {L.}~\bibnamefont
  {Strigari}}, \bibinfo {author} {\bibfnamefont {M.}~\bibnamefont
  {Kaplinghat}}, \bibinfo {author} {\bibfnamefont {G.}~\bibnamefont
  {Steigman}}, \ and\ \bibinfo {author} {\bibfnamefont {T.}~\bibnamefont
  {Walker}},\ }\href@noop {} {\bibfield  {journal} {\bibinfo  {journal}
  {Journal of Cosmology and Astroparticle Physics}\ }\textbf {\bibinfo {volume}
  {0403:007}} (\bibinfo {year} {2004})}\BibitemShut {NoStop}%
\bibitem [{\citenamefont {{Fukuda et. al.}}(2003)}]{sknim}%
  \BibitemOpen
  \bibfield  {author} {\bibinfo {author} {\bibnamefont {{Fukuda et. al.}}}
  (\bibinfo {collaboration} {{The Super-Kamiokande Collaboration}}),\
  }\href@noop {} {\bibfield  {journal} {\bibinfo  {journal} {Nucl. Instrum.
  Meth.}\ }\textbf {\bibinfo {volume} {A501}},\ \bibinfo {pages} {418}
  (\bibinfo {year} {2003})}\BibitemShut {NoStop}%
\bibitem [{\citenamefont {{M. Nakahata et al.}}(1999)}]{linacpaper}%
  \BibitemOpen
  \bibfield  {author} {\bibinfo {author} {\bibnamefont {{M. Nakahata et al.}}}
  (\bibinfo {collaboration} {{The Super-Kamiokande Collaboration}}),\
  }\href@noop {} {\bibfield  {journal} {\bibinfo  {journal} {Nucl. Instr.
  Meth.}\ }\textbf {\bibinfo {volume} {A421}},\ \bibinfo {pages} {113}
  (\bibinfo {year} {1999})}\BibitemShut {NoStop}%
\bibitem [{\citenamefont {Blaufuss}\ \emph {et~al.}(2001)\citenamefont
  {Blaufuss} \emph {et~al.}}]{n16paper}%
  \BibitemOpen
  \bibfield  {author} {\bibinfo {author} {\bibfnamefont {E.}~\bibnamefont
  {Blaufuss}} \emph {et~al.} (\bibinfo {collaboration} {Super-Kamiokande
  Collaboration}),\ }\href {\doibase 10.1016/S0168-9002(00)00900-1} {\bibfield
  {journal} {\bibinfo  {journal} {Nucl.Instrum.Meth.}\ }\textbf {\bibinfo
  {volume} {A458}},\ \bibinfo {pages} {638} (\bibinfo {year} {2001})},\ \Eprint
  {http://arxiv.org/abs/hep-ex/0005014} {arXiv:hep-ex/0005014 [hep-ex]}
  \BibitemShut {NoStop}%
\bibitem [{\citenamefont {Yanagisawa}\ and\ \citenamefont
  {Kato}(2001)}]{accident}%
  \BibitemOpen
  \bibfield  {author} {\bibinfo {author} {\bibfnamefont {C.}~\bibnamefont
  {Yanagisawa}}\ and\ \bibinfo {author} {\bibfnamefont {T.}~\bibnamefont
  {Kato}},\ }\href
  {http://www-sk.icrr.u-tokyo.ac.jp/cause-committee/1st/report-nov22e.pdf}
  {\emph {\bibinfo {title} {Report on the {Super-Kamiokande} Accident}}},\
  \bibinfo {type} {Tech. Rep.}\ (\bibinfo  {institution} {Stony Brook
  University},\ \bibinfo {year} {2001})\BibitemShut {NoStop}%
\bibitem [{\citenamefont {Nishino}\ \emph {et~al.}(2007)\citenamefont {Nishino}
  \emph {et~al.}}]{newdaq}%
  \BibitemOpen
  \bibfield  {author} {\bibinfo {author} {\bibfnamefont {H.}~\bibnamefont
  {Nishino}} \emph {et~al.}\ }(\bibinfo {year} {2007})\ pp.\ \bibinfo {pages}
  {127 --132}\BibitemShut {NoStop}%
\bibitem [{\citenamefont {Yamada}\ \emph {et~al.}(2008)\citenamefont {Yamada},
  \citenamefont {Obayashi}, \citenamefont {Shiozawa},\ and\ \citenamefont
  {Hayato}}]{newdaq2}%
  \BibitemOpen
  \bibfield  {author} {\bibinfo {author} {\bibfnamefont {S.}~\bibnamefont
  {Yamada}}, \bibinfo {author} {\bibfnamefont {Y.}~\bibnamefont {Obayashi}},
  \bibinfo {author} {\bibfnamefont {M.}~\bibnamefont {Shiozawa}}, \ and\
  \bibinfo {author} {\bibfnamefont {Y.}~\bibnamefont {Hayato}},\ }in\ \href
  {\doibase 10.1109/NSSMIC.2008.4774675} {\emph {\bibinfo {booktitle} {Nuclear
  Science Symposium Conference Record, 2008. NSS '08. IEEE}}}\ (\bibinfo {year}
  {2008})\ pp.\ \bibinfo {pages} {1387 --1390}\BibitemShut {NoStop}%
\bibitem [{\citenamefont {Strumia}\ and\ \citenamefont
  {Vissani}(2003)}]{strumia}%
  \BibitemOpen
  \bibfield  {author} {\bibinfo {author} {\bibfnamefont {A.}~\bibnamefont
  {Strumia}}\ and\ \bibinfo {author} {\bibfnamefont {F.}~\bibnamefont
  {Vissani}},\ }\href {\doibase 10.1016/S0370-2693(03)00616-6} {\bibfield
  {journal} {\bibinfo  {journal} {Phys. Lett.}\ }\textbf {\bibinfo {volume}
  {B564}},\ \bibinfo {pages} {42} (\bibinfo {year} {2003})},\ \Eprint
  {http://arxiv.org/abs/astro-ph/0302055} {arXiv:astro-ph/0302055} \BibitemShut
  {NoStop}%
\bibitem [{\citenamefont {Vogel}\ and\ \citenamefont
  {Beacom}(1999)}]{beacomvogel}%
  \BibitemOpen
  \bibfield  {author} {\bibinfo {author} {\bibfnamefont {P.}~\bibnamefont
  {Vogel}}\ and\ \bibinfo {author} {\bibfnamefont {J.~F.}\ \bibnamefont
  {Beacom}},\ }\href {\doibase 10.1103/PhysRevD.60.053003} {\bibfield
  {journal} {\bibinfo  {journal} {Phys. Rev. D}\ }\textbf {\bibinfo {volume}
  {60}},\ \bibinfo {pages} {053003} (\bibinfo {year} {1999})}\BibitemShut
  {NoStop}%
\bibitem [{\citenamefont {{J. Hosaka et al.}}(2006)}]{sk1solar}%
  \BibitemOpen
  \bibfield  {author} {\bibinfo {author} {\bibnamefont {{J. Hosaka et al.}}}
  (\bibinfo {collaboration} {{The Super-Kamiokande Collaboration}}),\
  }\href@noop {} {\bibfield  {journal} {\bibinfo  {journal} {Phys. Rev. D}\
  }\textbf {\bibinfo {volume} {73, 112001}} (\bibinfo {year}
  {2006})}\BibitemShut {NoStop}%
\bibitem [{\citenamefont {{J .P. Cravens et al.}}(2008)}]{sk2solar}%
  \BibitemOpen
  \bibfield  {author} {\bibinfo {author} {\bibnamefont {{J .P. Cravens et
  al.}}} (\bibinfo {collaboration} {{The Super-Kamiokande Collaboration}}),\
  }\href@noop {} {\bibfield  {journal} {\bibinfo  {journal} {Phys. Rev. D}\
  }\textbf {\bibinfo {volume} {78, 032002}} (\bibinfo {year}
  {2008})}\BibitemShut {NoStop}%
\bibitem [{\citenamefont {Ashie}\ \emph {et~al.}(2005)\citenamefont {Ashie}
  \emph {et~al.}}]{sk1oscpar}%
  \BibitemOpen
  \bibfield  {author} {\bibinfo {author} {\bibfnamefont {Y.}~\bibnamefont
  {Ashie}} \emph {et~al.} (\bibinfo {collaboration} {Super-Kamiokande}),\
  }\href {\doibase 10.1103/PhysRevD.71.112005} {\bibfield  {journal} {\bibinfo
  {journal} {Phys. Rev.}\ }\textbf {\bibinfo {volume} {D71}},\ \bibinfo {pages}
  {112005} (\bibinfo {year} {2005})},\ \Eprint
  {http://arxiv.org/abs/hep-ex/0501064} {arXiv:hep-ex/0501064} \BibitemShut
  {NoStop}%
\bibitem [{\citenamefont {Hayato}(2002)}]{neut}%
  \BibitemOpen
  \bibfield  {author} {\bibinfo {author} {\bibfnamefont {Y.}~\bibnamefont
  {Hayato}},\ }\href {\doibase DOI: 10.1016/S0920-5632(02)01759-0} {\bibfield
  {journal} {\bibinfo  {journal} {Nuclear Physics B - Proceedings Supplements}\
  }\textbf {\bibinfo {volume} {112}},\ \bibinfo {pages} {171 } (\bibinfo {year}
  {2002})}\BibitemShut {NoStop}%
\bibitem [{\citenamefont {Jegerlehner}\ \emph {et~al.}(1996)\citenamefont
  {Jegerlehner}, \citenamefont {Neubig},\ and\ \citenamefont
  {Raffelt}}]{raffelt}%
  \BibitemOpen
  \bibfield  {author} {\bibinfo {author} {\bibfnamefont {B.}~\bibnamefont
  {Jegerlehner}}, \bibinfo {author} {\bibfnamefont {F.}~\bibnamefont {Neubig}},
  \ and\ \bibinfo {author} {\bibfnamefont {G.}~\bibnamefont {Raffelt}},\ }\href
  {\doibase 10.1103/PhysRevD.54.1194} {\bibfield  {journal} {\bibinfo
  {journal} {Phys.Rev.}\ }\textbf {\bibinfo {volume} {D54}},\ \bibinfo {pages}
  {1194} (\bibinfo {year} {1996})},\ \Eprint
  {http://arxiv.org/abs/astro-ph/9601111} {arXiv:astro-ph/9601111 [astro-ph]}
  \BibitemShut {NoStop}%
\bibitem [{\citenamefont {Beacom}(2010)}]{Beacom:2010kk}%
  \BibitemOpen
  \bibfield  {author} {\bibinfo {author} {\bibfnamefont {J.~F.}\ \bibnamefont
  {Beacom}},\ }\href {\doibase 10.1146/annurev.nucl.010909.083331} {\bibfield
  {journal} {\bibinfo  {journal} {Ann. Rev. Nucl. Part. Sci.}\ }\textbf
  {\bibinfo {volume} {60}},\ \bibinfo {pages} {439} (\bibinfo {year} {2010})},\
  \Eprint {http://arxiv.org/abs/1004.3311} {arXiv:1004.3311 [astro-ph.HE]}
  \BibitemShut {NoStop}%
\bibitem [{\citenamefont {Y\"uksel}\ \emph {et~al.}(2006)\citenamefont
  {Y\"uksel}, \citenamefont {Ando},\ and\ \citenamefont {Beacom}}]{yukselando}%
  \BibitemOpen
  \bibfield  {author} {\bibinfo {author} {\bibfnamefont {H.}~\bibnamefont
  {Y\"uksel}}, \bibinfo {author} {\bibfnamefont {S.}~\bibnamefont {Ando}}, \
  and\ \bibinfo {author} {\bibfnamefont {J.~F.}\ \bibnamefont {Beacom}},\
  }\href {\doibase 10.1103/PhysRevC.74.015803} {\bibfield  {journal} {\bibinfo
  {journal} {Phys. Rev. C}\ }\textbf {\bibinfo {volume} {74}},\ \bibinfo
  {pages} {015803} (\bibinfo {year} {2006})}\BibitemShut {NoStop}%
\bibitem [{\citenamefont {Beacom}\ and\ \citenamefont
  {Vagins}(2004)}]{gadzooks}%
  \BibitemOpen
  \bibfield  {author} {\bibinfo {author} {\bibfnamefont {J.~F.}\ \bibnamefont
  {Beacom}}\ and\ \bibinfo {author} {\bibfnamefont {M.~R.}\ \bibnamefont
  {Vagins}},\ }\href {\doibase 10.1103/PhysRevLett.93.171101} {\bibfield
  {journal} {\bibinfo  {journal} {Phys. Rev. Lett.}\ }\textbf {\bibinfo
  {volume} {93}},\ \bibinfo {pages} {171101} (\bibinfo {year}
  {2004})}\BibitemShut {NoStop}%
\end{thebibliography}%

\end{document}